\documentclass[fleqn,usenatbib]{mnras}

\usepackage{newtxtext,newtxmath}

\usepackage[T1]{fontenc}
\usepackage{ae,aecompl}
\usepackage[normalem]{ulem}

\newcommand{\orcid}[1]{\href{https://orcid.org/#1}{\includegraphics[width=10pt]{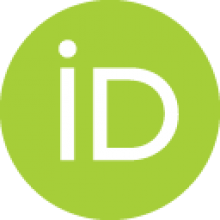}}}

\defcitealias{2019MNRAS.484.5702M}{Paper~I}
\defcitealias{2020MNRAS.498..205S}{Paper~II}
\defcitealias{2021A&A...647L...9D}{Paper~III}
\defcitealias{2022MNRAS.512.4334D}{Paper~IV}

\usepackage{graphicx}	
\usepackage{amsmath}	
\usepackage{multirow}
\usepackage{booktabs}

\title[Assembly history of the SMC Wing/Bridge region]
{The VISCACHA survey -- VII.
Assembly history of
the Magellanic Bridge and SMC Wing from star clusters}

\author[R.~A.~P. Oliveira et al.]
{R.~A.~P. Oliveira \orcid{0000-0002-4778-9243},$^{1}$\thanks{E-mail: \href{mailto:rap.oliveira@usp.br}{rap.oliveira@usp.br}}
F.~F.~S. Maia \orcid{0000-0002-2569-4032},$^{2}$
B. Barbuy \orcid{0000-0001-9264-4417},$^{1}$
B. Dias \orcid{0000-0003-4254-7111},$^{3}$
J.~F.~C. Santos Jr. \orcid{0000-0003-1794-6356},$^{4}$\newauthor
S.~O. Souza \orcid{0000-0001-8052-969X},$^{5,1}$
L.~O. Kerber \orcid{0000-0002-7435-8748},$^{6}$
E. Bica,$^{7}$
D. Sanmartim \orcid{0000-0002-9238-9521},$^{8,9}$
B. Quint,$^{9}$
L. Fraga,$^{10}$\newauthor
T. Armond,$^{11}$
D. Minniti,$^{12,13}$
M. C. Parisi,$^{14,15,16}$
O. J. Katime Santrich \orcid{0000-0003-0439-2331},$^{6}$
M. S. Angelo,$^{17}$ \newauthor
A. P\'erez-Villegas \orcid{0000-0002-5974-3998}$^{18}$
and B. J. De B\'ortoli$^{19,14,20}$
\\
$^{1}$Universidade de S\~ao Paulo, IAG, Rua do Mat\~ao 1226, Cidade Universit\'aria, S\~ao Paulo 05508-090, Brazil\\
$^{2}$Universidade Federal do Rio de Janeiro, Av. Athos da Silveira, 149, Cidade Universit\'aria, Rio de Janeiro 21941-909, Brazil\\
$^{3}$Instituto de Alta Investigaci\'on, Sede Esmeralda, Universidad de Tarapac\'a, Av. Luis Emilio Recabarren 2477, Iquique, Chile\\
$^{4}$Departamento de F\'isica, ICEx – UFMG, Av. Ant\^onio Carlos 6627, Belo Horizonte 31270-901, Brazil\\
$^{5}$Leibniz-Institut f\"ur Astrophysik Potsdam (AIP), An der Sternwarte 16, Potsdam D-14482, Germany\\
$^{6}$Universidade Estadual de Santa Cruz (UESC), Departamento de Ci\^encias Exatas, Rodovia Jorge Amado km 16, 45662-900 Ilh\'eus, Brazil\\
$^{7}$Universidade Federal do Rio Grande do Sul, Departamento de Astronomia, CP 15051, Porto Alegre 91501-970, Brazil\\
$^{8}$Carnegie Observatories, Las Campanas Observatory, Casilla 601, La Serena, Chile\\
$^{9}$Rubin Observatory Project Office, 950 N. Cherry Ave., Tucson, AZ 85719, USA\\
$^{10}$Laborat\'orio Nacional de Astrof\'isica, Rua Estados Unidos 154, Itajub\'a 37504-364, Brazil\\
$^{11}$Universidade Federal de S\~ao Jo\~ao del-Rei, Departamento de Estat\'istica, F\'isica e Matem\'atica, Campus Alto Paraopeba, Rod.: MG 443, Km\,7, Ouro Branco -\\ MG 36420-000, Brazil\\
$^{12}$Instituto de Astrof\'isica, Facultad de Ciencias Exactas, Universidad Andres Bello, Av. Fern\'andez Concha 700, Santiago, Chile\\
$^{13}$Vatican Observatory, V00120 Vatican City State, Italy\\
$^{14}$Consejo Nacional de Investigaciones Cient\'ificas y T\'ecnicas, Godoy Cruz 2290, C1425FQB, Ciudad Aut\'onoma de Buenos Aires, Argentina\\
$^{15}$Observatorio Astron\'omico, Universidad Nacional de C\'ordoba, Laprida 854, X5000BGR, C\'ordoba, Argentina\\
$^{16}$Instituto de Astronom\'ia Te\'orica y Experimental (CONICET-UNC), Laprida 854, X5000BGR, C\'ordoba, Argentina\\
$^{17}$Centro Federal de Educa\c c\~ao Tecnol\'ogica de Minas Gerais, Av. Monsenhor Luiz de Gonzaga, 103, 37250-000 Nepomuceno, MG, Brazil\\
$^{18}$Instituto de Astronom\'ia, Universidad Nacional Aut\'onoma de M\'exico, Apartado Postal 106, C. P. 22800, Ensenada, B. C., Mexico\\
$^{19}$Instituto de Astrofísica de La Plata, CONICET–UNLP, Paseo del Bosque S/N, B1900FWA La Plata, Argentina\\
$^{20}$Facultad de Ciencias Astron\'omicas y Geof\'isicas, UNLP, Paseo del Bosque S/N, B1900FWA La Plata, Argentina
}

\date{Accepted 2023 June 7. Received 2023 June 6; in original form 2023 February 7}

\pubyear{2023}

\begin{document}
\label{firstpage}
\pagerange{\pageref{firstpage}--\pageref{lastpage}}
\maketitle

\begin{abstract} 
The formation scenario of the Magellanic Bridge during
an encounter
between the Large and Small Magellanic Clouds $\sim200$\,Myr ago, as proposed by
\textit{N}-body
models,
would be imprinted in the chemical enrichment and kinematics of its stars, 
and sites of ongoing star formation along its extension. 
We present an analysis of 33 
Bridge star clusters using photometry
obtained
with
the
SOAR 4-m telescope
equipped with
adaptive optics for the VISCACHA survey.
We performed a membership selection and derived self-consistent ages, metallicities, distances and
reddening values
via statistical isochrone fitting, as well as tidal radii and integrated masses from structure analysis. 
Two groups
are clearly detected: 13 well-studied clusters older than the Bridge,
with $0.5-6.8$\,Gyr and
$\rm{[Fe/H]}<-0.6$\,dex;
and 15 clusters with $< 200$\,Myr and
$\rm{[Fe/H]}>-0.5$\,dex,
probably formed in-situ. 
The old clusters follow the overall age and metallicity gradients of the SMC, whereas the younger ones are uniformly distributed along the Bridge. 
The
main results are as follows:
$(i)$
we derive
ages and metallicities
for the first time for 9 and 18 clusters, respectively; $(ii)$
we detect
two metallicity dips in the age-metallicity relation of the Bridge at
$\sim 200$\,Myr
and $1.5$\,Gyr ago
for the first time,
possibly
chemical signatures of the formation of the Bridge and Magellanic Stream; $(iii)$ we estimate a minimum stellar mass for the Bridge of $3-5 \times 10^5 M_\odot$; $(iv)$ we confirm that
all the young Bridge clusters
at
$\rm{RA} < 3^h$ are
metal-rich
with $\rm{[Fe/H]} \sim -0.4$\,dex.
\end{abstract}




\section{Introduction}

The Magellanic System consists of the Large and Small Magellanic Clouds (LMC and SMC, or jointly MCs),
the Magellanic Stream, Bridge and Leading Arm. The Bridge contains mostly HI gas, but also a stellar population with a mass of $1.5\times10^4M_\odot$ \citep{2007ApJ...658..345H}. In particular, the existence of the Stream and Bridge provides evidence of tidal interactions between the MCs, and studying their metallicity content and stellar ages can enlighten our knowledge on the Bridge formation.

A few characteristics of the LMC and SMC structure are also evidence of their interaction process and formation of the Bridge.
The LMC
is
located at a distance of $49.6\pm0.5$\,kpc \citep[][]{2019Natur.567..200P}
and
presents a flat disk morphology with a single spiral arm
and an asymmetric
stellar bar \citep[e.g.][]{2001AJ....122.1827V, 2020A&A...639L...3R}, as well as a warped outer stellar disk possibly connected to the Bridge \citep[e.g.][]{2018ApJ...866...90C, 2022A&A...666A.103S}.
The SMC is located at a distance of $62.4\pm0.8$\,kpc \citep{2020ApJ...904...13G} and, due to strong interactions with the LMC, presents an elongated, triaxial structure, with a line-of-sight depth of up to $14$\,kpc in the inner regions \citep{2012ApJ...744..128S} and $\sim 23$\,kpc in the eastern part \citep{2013ApJ...779..145N}.
The Bridge extent corresponds to the physical distance between the MCs ($\sim 20$\,kpc), with a $\sim 10$\,kpc depth,
as traced by Cepheid and RR Lyrae stars
\citep{2016AcA....66..149J, 2017AcA....67....1J, 2017MNRAS.472..808R, 2018MNRAS.473.3131M}.

A mean metallicity of
$\rm{[Fe/H]}=-0.37\pm0.15$\,dex
was obtained from
CaII triplet (CaT) lines
for 373 LMC red giants by  \citet{2005AJ....129.1465C}. 
The SMC has a mean metallicity of
$\rm{[Fe/H]}=-0.99\pm0.01$\,dex
\citep{2014MNRAS.442.1680D} and
$-0.9\pm0.2$\,dex
\citep{2016AJ....152...58P}, derived from cluster and field red giants also using CaT lines.
The metallicity gradients
in the inner regions of the LMC and 
SMC are relatively well known
(e.g. \citealt{2009AJ....138..517P, 2016AJ....152...58P, 2016MNRAS.455.1855C, 2018MNRAS.475.4279C, 2023arXiv230519460M}; Nidever et al. in prep.),
whereas the outer regions
are less explored and present much higher dispersion. For the SMC, for example, \citet{2014AJ....147...71P, 2015AJ....149..154P, 2022A&A...662A..75P} and \citet{2014MNRAS.442.1680D} detected
a decreasing trend in metallicity
up to a semi-major axis $a \sim 4.5^\circ$ and an inversion after that,
using cluster and field stars.
A similar trend was obtained in \citet{2020AJ....159...82B},
who made a compilation of star clusters with heterogeneous ages and metallicities from the literature.

The Bridge is closely related to the MCs,
sharing a common HI envelope with
them particularly as a continuity of the SMC Wing \citep[e.g.][]{2003ApJ...586..170P}.
Metallicity measurements of the gas from absorption lines resulted to be $\rm{[M/H]} = -1.0$ and
$-1.3$\,dex
for two massive stars in different sightlines \citep{2008ApJ...678..219L}, and
$-0.8$\,dex using
a background
quasar \citep{2009ApJ...695.1382M}. These values are closer to the SMC metallicity than to that of the LMC, suggesting that the Bridge may have been formed mostly from SMC material, even though radial velocity measurements indicate that LMC gas might be present as well
\citep{1986PASA....6..471M}.

The Bridge contains a blue, young stellar population 
detected by \citet{1985Natur.318..160I},
as well as intermediate-age ($1-3$\,Gyr) and old stellar populations ($> 3$\,Gyr) 
detected later
\citep{2013A&A...551A..78B, 2013ApJ...768..109N, 2017MNRAS.466.4711B}.
The young population
presents a higher concentration closer to the SMC Wing and another halfway between the MCs,
with a strong correlation with the distribution and kinematics of the HI gas \citep[e.g.][]{2014ApJ...795..108S, 2019ApJ...874...78Z, 2019ApJ...887..267M}.
On the other hand, the old population is uniformly distributed between the MCs, with a
large number
of RR Lyrae stars in the southern part \citep{2017MNRAS.466.4711B}.
Recent
proper motions from \textit{Hubble Space Telescope}, \textit{Gaia} and Visible and Infrared Survey Telescope for Astronomy (VISTA) showed
a flow motion
of both young and red clump stars in the Wing/Bridge pointing from the SMC towards the LMC \citep{2019ApJ...874...78Z, 2020A&A...641A.134S, 2021A&A...649A...7G}, with larger motions in regions of lower stellar density.
\citet{2020A&A...641A.134S} found that the Bridge is stretching as its borders close to the SMC and LMC are moving apart relative to the proper motion of its central region.

Based on different proper motion epochs \citep{2006ApJ...638..772K, 2006ApJ...652.1213K, 2013ApJ...764..161K} and initial conditions, two main $N$-body simulations attempt to reproduce the formation of the LMC-SMC pair, providing different scenarios:
the LMC captured the SMC $\sim 1.2$\,Gyr ago and they are in a bound orbit around the
Milky Way (MW) since then
\citep[often referred to as bound scenario;][]{2011MNRAS.413.2015D}, or it is an old interacting system formed about $5-6$\,Gyr ago in its first perigalactic passage, falling into the MW potential since $\sim2$ Gyr ago \citep[unbound scenario;][]{2007ApJ...668..949B, 2012MNRAS.421.2109B}.
Although more evidence points to the unbound scenario \citep[e.g.][]{2021Natur.592..534C},
the masses of the MW and LMC are the main sources of errors.
\citet{2020ApJ...893..121P} computed orbits of MW and LMC satellites adopting masses of
$1.0$ and $1.5\times10^{12}\,M_\odot$ for the MW, and $0.8$, $1.8$ and
$2.5\times10^{11}\,M_\odot$ for the LMC, showing the impact of this uncertainty on the
orbits of the system members. Even so, a study on age and metallicity of Bridge objects
might shed some light on its formation.
Some works focused on simulating the formation and filamentary structure of the Stream \citep[e.g.][]{2012ApJ...750...36D, 2015ApJ...813..110H}, reproducing well the HI observations.

Both scenarios are able to reproduce the large-scale gas structures (Stream, Leading Arm and gaseous Bridge), with the Bridge being formed during the most recent encounter between the MCs, with gas and stars pulled out from the SMC through tidal interactions and possibly dragged from the LMC to the Bridge \citep{2012MNRAS.421.2109B}. There is an uncertainty on the exact epoch of this encounter: $250$\,Myr \citep{2011MNRAS.413.2015D}, $200$\,Myr \citep{1996MNRAS.278..191G, 2012MNRAS.421.2109B}, $150$\,Myr \citep{2018ApJ...864...55Z} and $< 250$\,Myr with an impact parameter of $5-10$\,kpc \citep{2022ApJ...927..153C}.
This scenario of the Bridge formation
would imply a
gradient of increasing metallicity toward the LMC due to a minor contribution of its more metal-rich gas \citep{2012MNRAS.421.2109B},
and the presence of an older stellar population stripped from the SMC
amidst a young population formed in situ.
The recent encounter could also have triggered the formation of a ring-like structure in the LMC periphery \citep[also found in the distribution of star clusters, e.g.][]{2001AJ....122.1827V, 2008MNRAS.389..678B, 2018ApJ...869..125C} and of the warp along the northeast-southwest direction.

\begin{table}
	\caption{Log of observations, separated by semester. The seeing and full width at half maximum (FWHM) are given for the $V$ and $I$ filters, respectively. The observations were carried out with the SAMI imager, combining short and long exposures ($3\times400$\,s with $V$ and $3\times600$\,s with $I$).}
	\label{tab:obslog}
    \begin{center}
	\begin{tabular}{lcccc}
		\hline
		Cluster & Date & Airmass & Seeing$^\dagger$ & FWHM \\ 
		& & & (arcsec) & (arcsec) \\ 
		\hline
        K55 & 2016-09-28 & 1.47 & 0.8, 0.9 & 1.0, 0.8 \\ 
		
		K57 & 2016-09-28 & 1.55 & 0.7, 0.8 & 0.8, 0.7 \\ 

		HW71se & 2016-11-03 & 1.35 & 1.4, 1.3 & 1.3, 0.9 \\ 
		
		HW77 
        & 2016-11-05 
        & 1.36 & 0.9, 0.9 & 0.5, 0.3 \\ 
		
		BS187 
        & 2016-11-03 & 1.38 & 1.3, -- & 1.4, 1.1 \\ 
		
		BS198 & 2016-09-24 & 1.49 & 1.1, 1.3 & 1.1, 0.7 \\ 

        L113 & 2016-11-05 & 1.47 & 0.9, 1.0 & 0.6, 0.4 \\ 
		
		L114 & 2016-11-05 
        & 1.43 & 0.8, 0.9 & 0.5, 0.4 \\ 
		
		NGC\,796 & 2016-11-05 & 1.77 & -- , -- & 0.6, 0.4 \\ 
		
		\hline
		
		L92+L93 & 2017-10-22 & 1.45 & 1.0, 1.0 & 1.0, 0.9 \\ 
		L109 
        & 2017-10-22 & 1.60 & 0.8, 0.8 & 0.9, 0.6 \\ 
		HW86 & 2017-10-22 & 1.49 & 1.0, 0.9 & 0.9, 0.7 \\ 
		\hline
		HW55 
        & 2018-10-05 & 1.55 & 0.8, 0.7 & 1.1, 1.0 \\ 
		OGLB\,33 & 2018-12-12 & 1.39 & 0.8, 0.8 & 0.9, 0.6 \\ 
		\hline
		HW63
        & 2019-10-05 & 1.53 & $-$, $-$ & 0.9, 1.0 \\ 
		L91 & 2019-10-05 & 1.37 & $-$, $-$ & 0.7, 0.6 \\ 
		B147 & 2019-10-05 & 1.45 & $-$, $-$ & 0.8, 0.7 \\ 
		WG1 & 2019-10-05 & 1.52 & $-$, $-$ & 0.8, 0.7 \\ 
		WG13 & 2019-12-22 & 1.46 & 0.8, 0.7 & 0.9, 0.7 \\ 
		\hline
		BS245 & 2020-11-11 & 1.43 & 0.8, $-$ & 0.8, 0.5 \\ 
		HW75 & 2020-11-11 & 1.41 & 0.6, 0.5 & 0.6, 0.4 \\ 
		HW78 & 2020-11-13 & 1.39 & $-$, $-$ & 1.3, 1.1 \\ 
		HW81+HW82 & 2020-11-11 & 1.37 & $-$, $-$ & 0.7, 0.5 \\ 
		L101 & 2020-11-13 & 1.37 & 1.1, $-$ & 1.5, 1.2  \\ 
		L104 & 2020-11-12 & 1.40 & $-$, 0.4 & 0.7, 0.4 \\ 
		L107 & 2020-11-12 & 1.38 & 0.8, 0.6 & 0.6, 0.5 \\ 
		L110 & 2020-11-11 & 1.37 & $-$, 0.7 &  0.6, 0.4 \\ 
		\hline
		HW59 
        & 2021-11-07 & 1.44 & $-$, $-$ & 0.6, 0.6 \\ 
		ICA45 & 2021-11-09 & 1.39 & 0.6, $-$ & 0.7, 0.5 \\ 
		B165 & 2021-11-11 & 1.44 & 0.8, 0.9 & 1.2, 0.9 \\ 
		BS226 & 2021-11-11 & 1.43 & 0.9, $-$ & 0.9, 0.9 \\ 
    \hline
		
	\end{tabular}
    \end{center}
 
    \noindent
 
    \noindent
    
    \noindent 
    $^\dagger$ Those marked with $-$ could not be retrieved from the site seeing monitor. 
\end{table}

The stellar clusters located in the Magellanic System are important tracers of the star formation history and chemodynamical evolution of the LMC-SMC pair.
A catalog
of 2741 stellar clusters, associations and extended objects in the SMC and Bridge was reported by \citet[and references therein]{2020AJ....159...82B}, containing over 400 objects in the Wing/Bridge.
Most of the Wing/Bridge objects are younger
than 1\,Gyr \citep[e.g.][]{2015MNRAS.453.3190B}, but older objects are found as well, reaching $\sim 6$\,Gyr \citep[e.g.][]{2015MNRAS.450..552P}.

In the present work, we analysed 33 
clusters located in the Wing/Bridge, that were observed with the adaptive optics system of the 4-m SOuthern Astrophysical Research (SOAR) telescope within the VIsible Soar photometry of star Clusters in tApii and Coxi HuguA \citep[VISCACHA survey\footnote{\url{http://www.astro.iag.usp.br/~viscacha/}};][]{2019MNRAS.484.5702M}. We derived structural parameters from radial density profiles, as well as age, metallicity, distance and reddening from decontaminated colour-magnitude diagrams (CMDs), with the aim of verifying the
existence of
gradients.

This paper is structured as follows. Observations and reductions are described in Section~\ref{sec2}. Methods of analysis (radial density profile fitting, decontamination and isochrone fitting) are detailed in Section~\ref{sec3}. Results are presented in Section~\ref{sec4}. The age and metallicity gradients, followed by a thorough discussion about their implications to the dynamical models, are given in Section~\ref{sec5}. Conclusions are drawn in Section~\ref{sec6}.




\section{Observations and data reduction}
\label{sec2}

\begin{figure*}
    \centering 
    \includegraphics[width=0.83\textwidth]{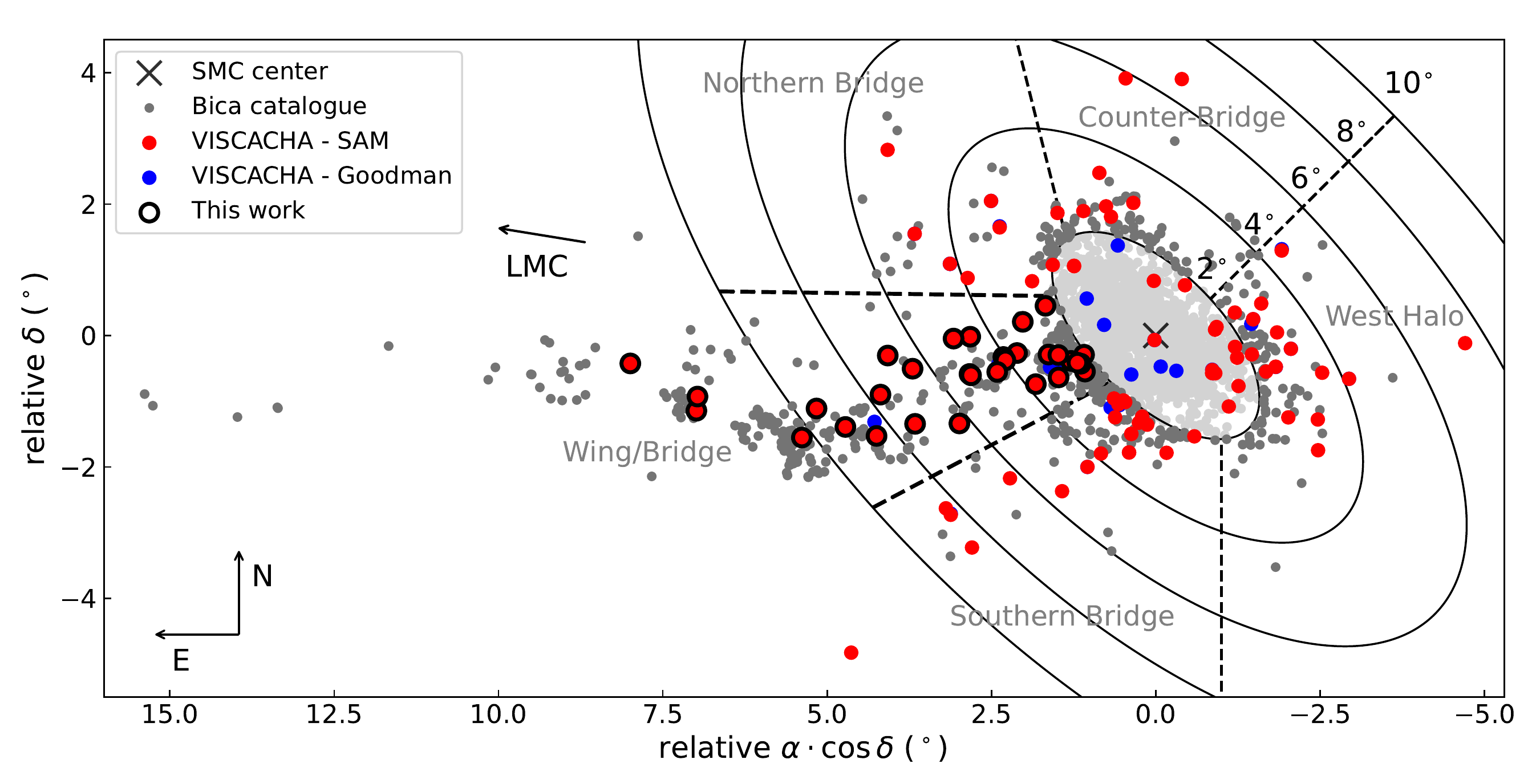}
    \caption{Projected distribution of the 2665 stellar clusters and associations from the \citet{2020AJ....159...82B} catalogue (grey dots), with coordinates relative to the SMC centre \citep[$0^{\rm h}52^{\rm m}45^{\rm s}, -72^\circ 49^\prime 43^{\prime\prime}$;][]{2001AJ....122..220C}.
    The arrow connects the direction of the SMC and LMC centres.
    The objects observed
    with SAM and Goodman are marked in red and blue respectively, and the present sample of 33 
    Bridge clusters are marked with a black border.}
    \label{fig:ellipses}
\end{figure*}

\begin{figure*}
    \centering
    \includegraphics[height=4.85cm]{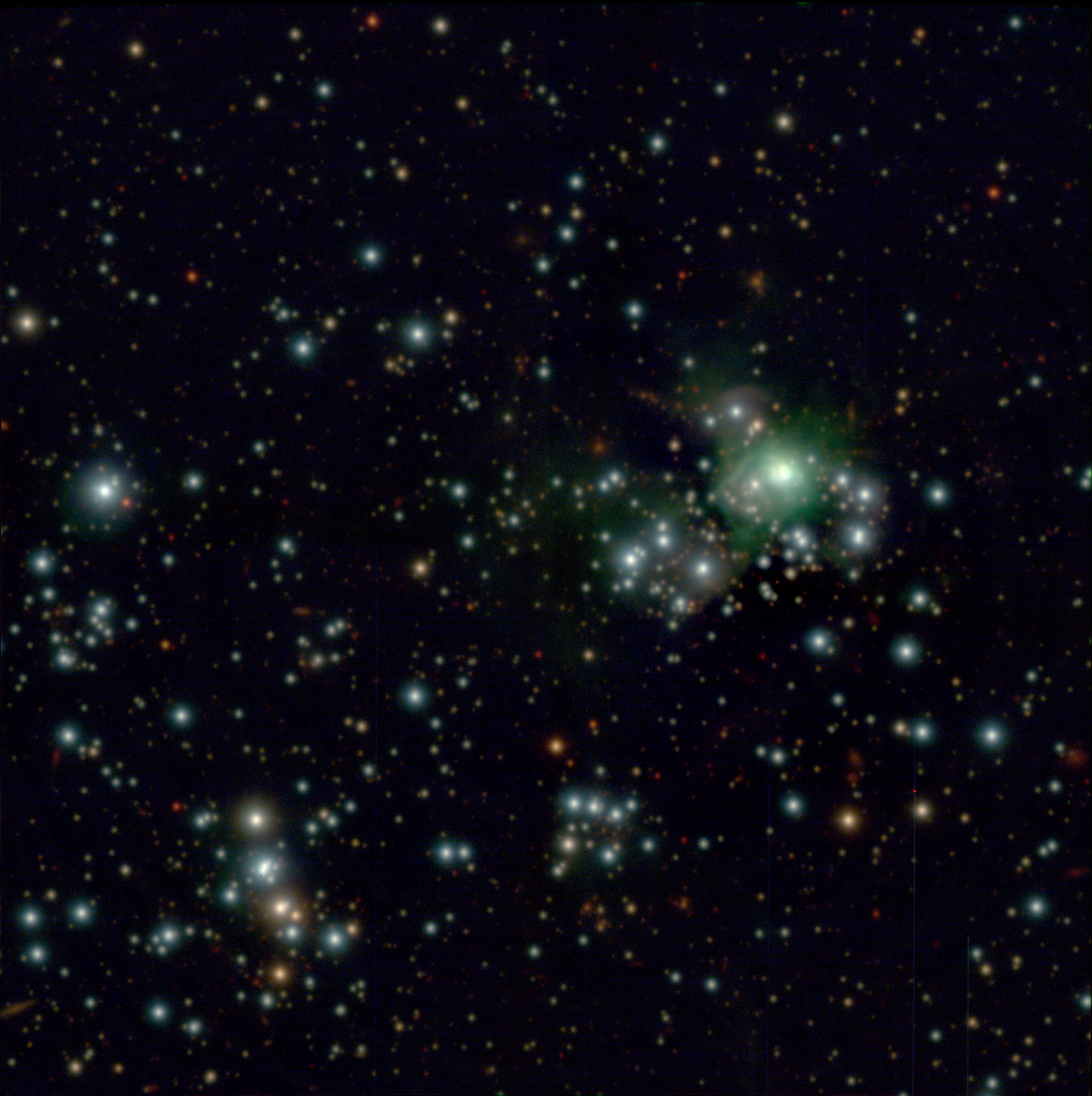}
    \hspace{0.5mm}
    \includegraphics[height=4.85cm]{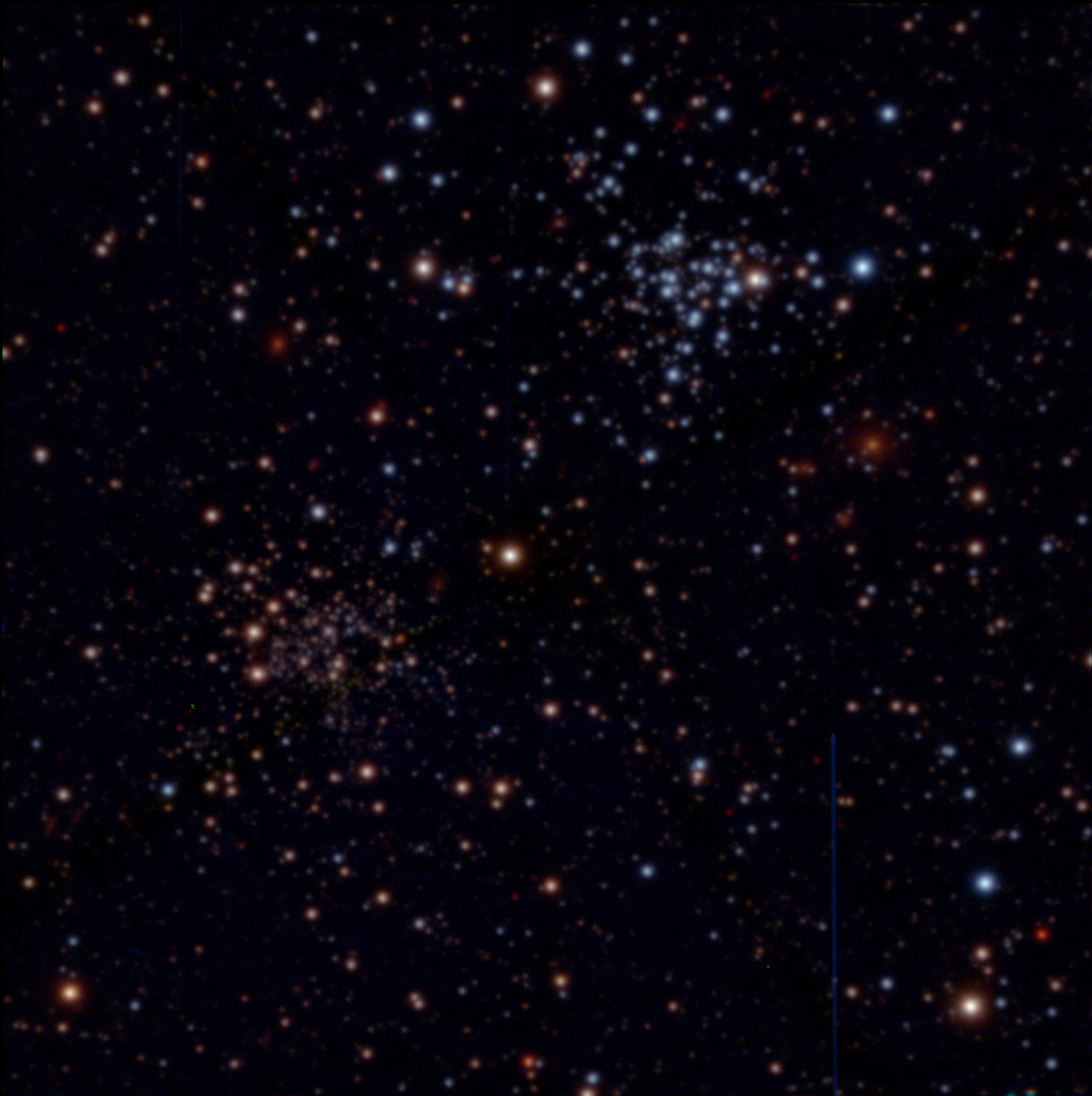}
    \hspace{0.5mm}
    \includegraphics[height=4.85cm]{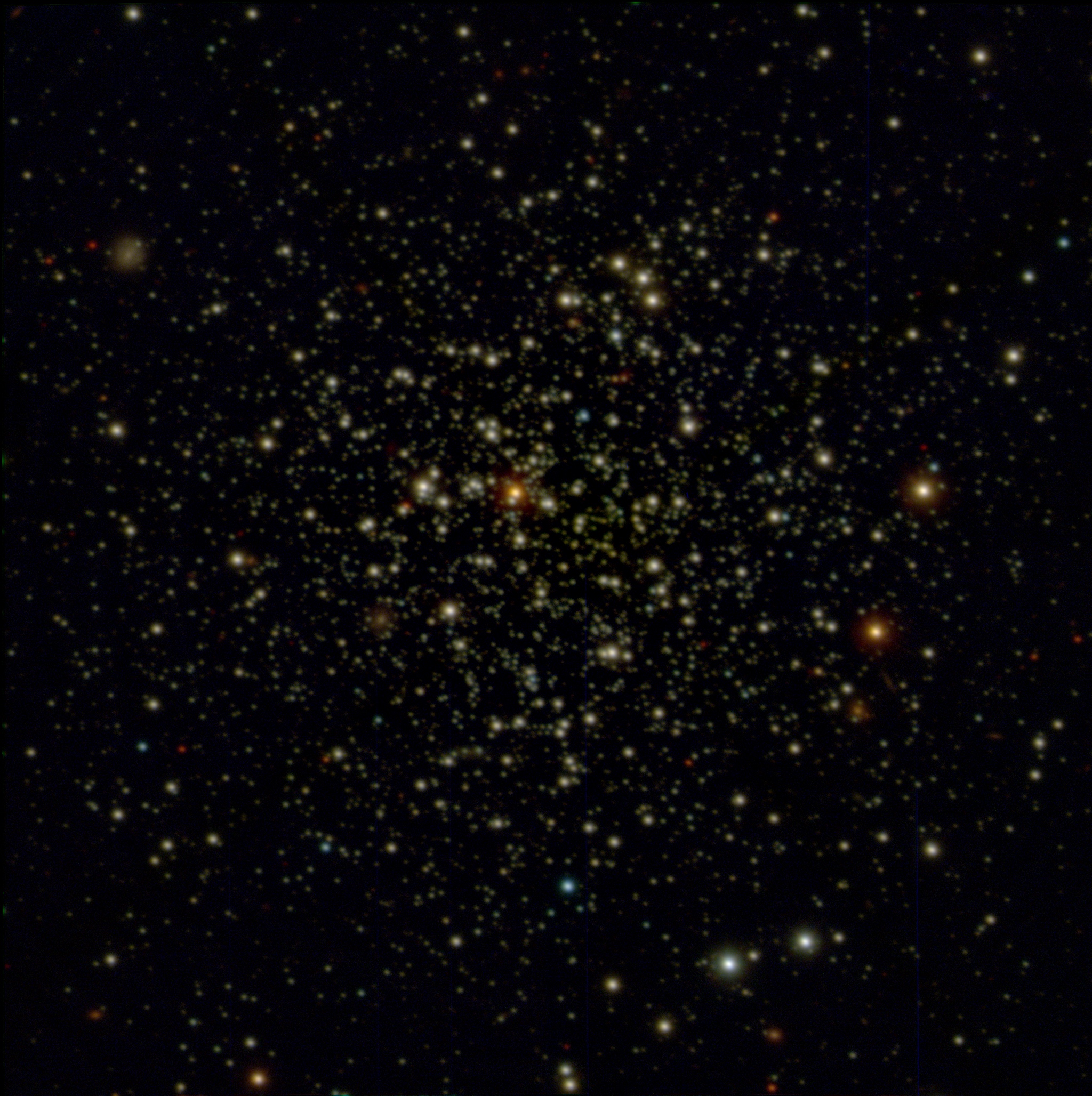}
    \caption{Colour composite images of HW81+HW82, L92+L93,
    and L113 respectively, obtained with SAMI ($3\times3$\,arcmin$^2$).
    North is up and East to the left.} 
    \label{fig:RGB_L113}
\end{figure*}

The observations were carried out through 2016B-2021B semesters, using the 4.1\,m SOAR (SOuthern Astrophysical Research) telescope, mounted with its ground-layer adaptive optics imager \citep[SAMI;][]{2013aoel.confE..12T,2016PASP..128l5003T} and using a 4K $\times$ 4K CCD detector with $2 \times 2$ binning, yielding a plate scale of $0.091$\,arcsec\,pixel$^{-1}$ and a field of view of $3.1\times 3.1\,\rm{arcmin}^2$. The cluster sample was observed in $V$ and $I$ filters with effective exposure times of 20 and 30 minutes (after co-addition), respectively, reaching
a signal-to-noise ratio of $\sim 10$ and final errors of $\sim 0.1$\,mag at
$V \sim24$\,mag
\citep[see][hereafter \citetalias{2019MNRAS.484.5702M}]{2019MNRAS.484.5702M}.
Short exposures were also taken to sample saturated stars. Table~\ref{tab:obslog} presents the log of observations for the sample clusters, containing the observation date, airmass, seeing and the measured image quality ($<1$\,arcsec in most cases). 

Figure~\ref{fig:ellipses} gives the projected distribution of the SMC and Bridge objects relative to the SMC centre. Concentric ellipses with semi-major axis from 2 to $10^\circ$ \citep[$b/a=1/2$, inclination of $45^\circ$;][]{2005A&A...440..111P} are used to obtain the projected cluster distance to the SMC centre.
The clusters observed in the VISCACHA survey are marked in red (SAMI) and blue (Goodman, backup instrument), whereas the present sample is indicated with black
border
circles.
Figure~\ref{fig:RGB_L113} shows the colour composite images obtained with the $V$ and $I$ filters for the pairs HW81+HW82 (two young clusters), L92+L93 (young vs. old cluster) and L113 (old, populous cluster). 

The data were processed for bias subtraction and division by skyflats in a standard way using the IRAF CCDRED package. Cosmic rays were removed from the images using the CRUTIL package. Astrometric calibration was performed with IRAF MSCCMATCH task using \textit{Gaia} Early Data Release 3 catalogues \citep{2021A&A...649A...1G} as astrometric reference, this way achieving typical RMS residuals inferior to 0.1\,arcsec on source positions. The images World Coordinate Systems were then used to register them to a common frame and co-add them into the final mosaics. PSF photometry was carried out using a modified version of the \texttt{STARFINDER} code \citep{2000A&AS..147..335D}. For the photometric calibrations, we employed \citet{2000PASP..112..925S} and MCPS \citep[Magellanic Clouds Photometric Survey;][]{2002AJ....123..855Z} fields, when available. Further details are given in
\citetalias{2019MNRAS.484.5702M}.

\begin{table}
	\centering
	\caption{Star cluster parameters from the literature, obtained either from photometric or spectroscopic data.}
	\label{tab:literature}
	\begin{tabular}{lcc}
		\hline
        \multirow{2}{*}{Cluster} & Age & $\rm{[Fe/H]}$ \\
		& (Gyr) & (dex) \\ 
		\hline

        HW55 & $1.00\pm1.15^1$, $2.5\pm0.7^{2}$ & $-0.40\pm0.22^3$ \\
        & $1.58\pm2.24^3$ \\ 

        K55 & $0.25 \pm 0.12^1$, $0.28\pm0.03^4$ & $-0.58\pm0.33^3$ \\
        & $0.63\pm0.07^3$ \\ 
        
        K57 & $0.45\pm0.31^1$, $0.45\pm0.05^4$ & $-0.48\pm0.26^3$ \\
        & $0.56\pm0.06^3$ \\ 

        HW59 & $6.7\pm1.1^5$, $7.9\pm5.5^3$ & $-0.88\pm1.3^3$ \\ 

        HW63 & $0.45\pm0.31^1$, $5.4\pm1.0^5$ & $-0.70\pm0.43^3$ \\
        & $3.55\pm0.49^3$ \\ 

        L92 & $0.13\pm0.09^1$ & --- \\ 
        
        L93 & $1.00\pm0.69^1$ & --- \\ 
        
        L91 & $0.79\pm0.55^1$, $4.3\pm1.0^5$ 
        & $-0.88\pm0.65^3$, $-0.90\pm0.06^{6}$ \\
        & $4.0\pm0.55^3$ \\ 

        B147 & $0.13\pm0.06^1$ & --- \\ 

        HW71se & $< 0.10^1$, $0.06^{+0.10}_{-0.02}$ (7) & --- \\ 

        HW75 & $0.16\pm0.11^1$, $0.20\pm0.05^8$ & --- \\ 

        HW77 & $1.41\pm0.32^8$ & --- \\ 


        HW81 & $0.010\pm0.002^8$ & --- \\ 
        
        HW82 & $0.06\pm0.01^8$ & --- \\ 

        BS187 & $2.00\pm0.46^8$ & --- \\ 


        L109 & $2.5\pm0.6^9$, $4.0\pm0.9^8$ 
        & $-0.88\pm0.65^3$ \\ 
        & $5.0\pm2.3^3$ \\ 

        L110 & $6.4\pm1.1^{10}$, $6.3\pm1.5^8$ & $-1.03\pm0.05^{11}$, $-0.88\pm0.65^3$ \\ 
        & $5.0\pm0.8^3$ \\ 

        HW86 & $1.7\pm0.2^{12}$, $1.41\pm0.32^8$ & $-0.61\pm0.06^{11}$ \\ 

        L113 & $5.3\pm1.0^{10}$, $3.55\pm0.41^3$ & $-1.12\pm0.12^{14}$, $-1.03\pm0.04^{15}$ \\
        & $3.75\pm0.30^{13}$ & $-0.88\pm0.65^3$ \\ 

        L114 & $0.14\pm0.03^{16}$, $0.16\pm0.07^3$ & $-0.10\pm0.11^3$ \\ 
        
        NGC796 & $0.11\pm0.03^{16}$, $0.04\pm0.02^{17}$ & $-0.31\pm0.10^{19}$ \\ 
        & $0.02\pm0.01^{18}$, $0.04\pm0.02^{19}$ \\
        
        WG13 & $0.13\pm0.07^{17}$ & $-0.2\pm0.26^{17}$ \\

        BS226 & $0.89\pm0.31^{17}$ & $-0.88\pm 0.43^{17}$ \\ 
        
        BS245 & $0.10\pm0.06^{17}$ & $-0.28\pm0.33^{17}$ \\
        
    \hline
	\end{tabular}
    \begin{list}{\textbf{References.}}
        \item (1) \citet
        {2010A&A...517A..50G};
        (2) \citet
        {2011MNRAS.418L..69P}; 
        (3) \citet
        {2017A&A...602A..89P};
        (4) \citet
        {2014MNRAS.437.2005M};
        (5) \citet
        {2011MNRAS.416L..89P}; 
        (6) \citet
        {2022A&A...664A.168D};
        (7) \citet
        {2005AJ....129.2701R};
        (8) \citet
        {2015MNRAS.450..552P};
        (9) \citet
        {2011MNRAS.417.1559P}; 
        (10) \citet{2007MNRAS.381L..84P};
        (11) \citet{2009AJ....138..517P};
        (12) \citet{2014AJ....147...71P};
        (13) \citet{2021A&A...647A.135N};
        (14) \citet{1998AJ....115.1934D};
        (15) \citet{2015AJ....149..154P};
        (16) \citet{2007MNRAS.382.1203P};
        (17) \citet{2015MNRAS.453.3190B};
        (18) \citet{2018ApJ...857..132K};
        (19) \citet[\citetalias{2019MNRAS.484.5702M}]{2019MNRAS.484.5702M}.
        
    \end{list}
\end{table}

A total of 33 
clusters were observed inside 31 fields.
HW81 and HW82 were observed in a same field, and there is apparently a third structure named HW81w by \citet{2020AJ....159...82B} that was not retrieved. HW71se also has a nearby cluster named HW71nw,
which we could not detect.
Six associations observed with Goodman
will be analysed in a future work: 
NGC456 and NGC460 (2021B, 1 night); ASS67, WG5, ASS65 and ASS66 (2020B, 2 nights).
The VISCACHA sample contains five clusters in the LMC western side close to the Bridge, observed in 2017B, but
we will defer the LMC-Bridge connection to a future work, after
the observations of the clusters in this region are completed.





\begin{figure}
    \centering
    \includegraphics[width=\columnwidth]{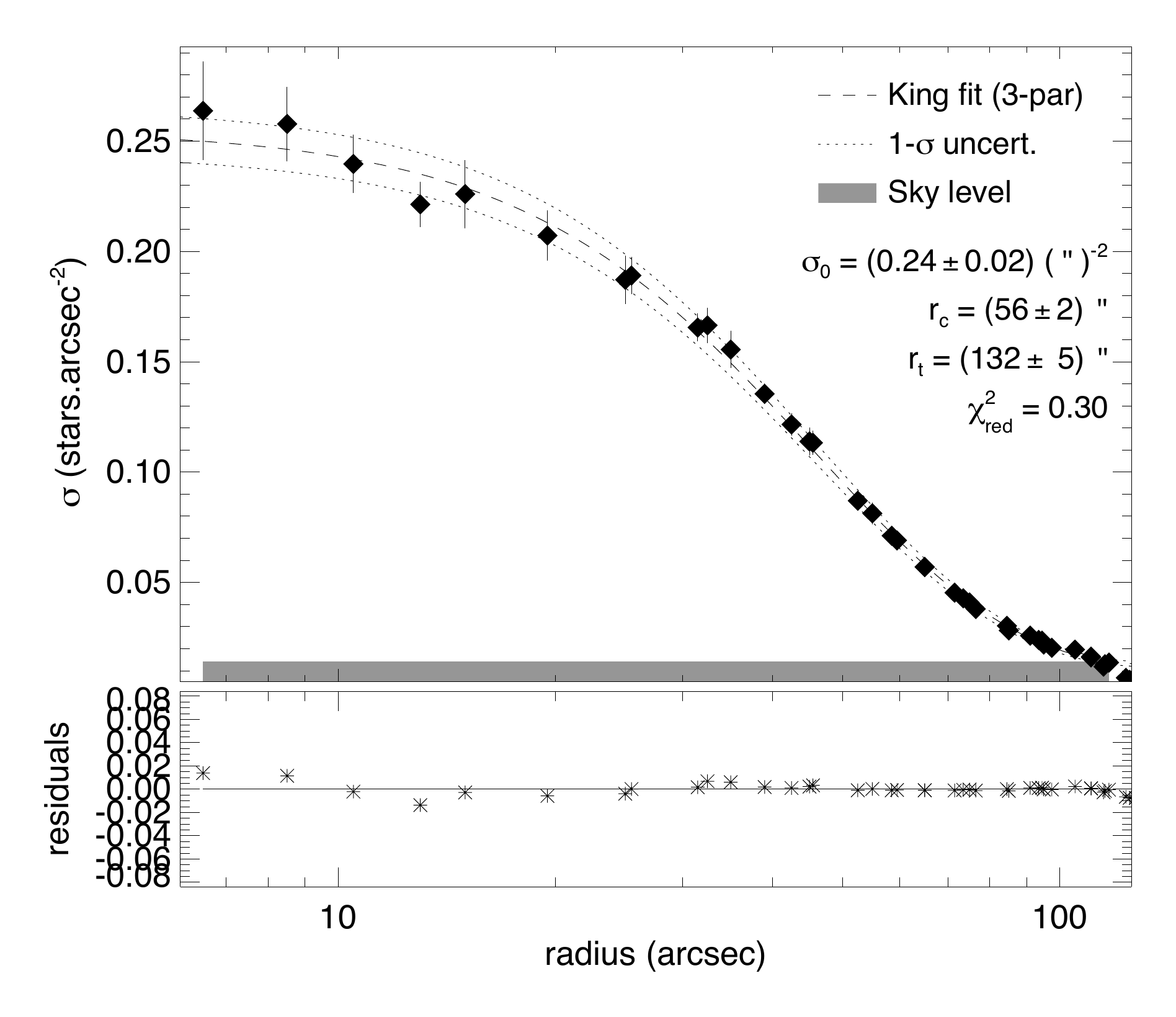}
    \caption{Example of the density profile fitting for the cluster L113, showing the best King model (dashed line) and $1\sigma$ uncertainties (dotted lines) derived. Error bars correspond to Poisson uncertainties. The derived structural parameters are given in the legend.}
    \label{fig:rdp}
\end{figure}

\begin{figure*}
    \centering
    \includegraphics[trim={0 0.4cm 0 0},clip,width=0.9\textwidth]{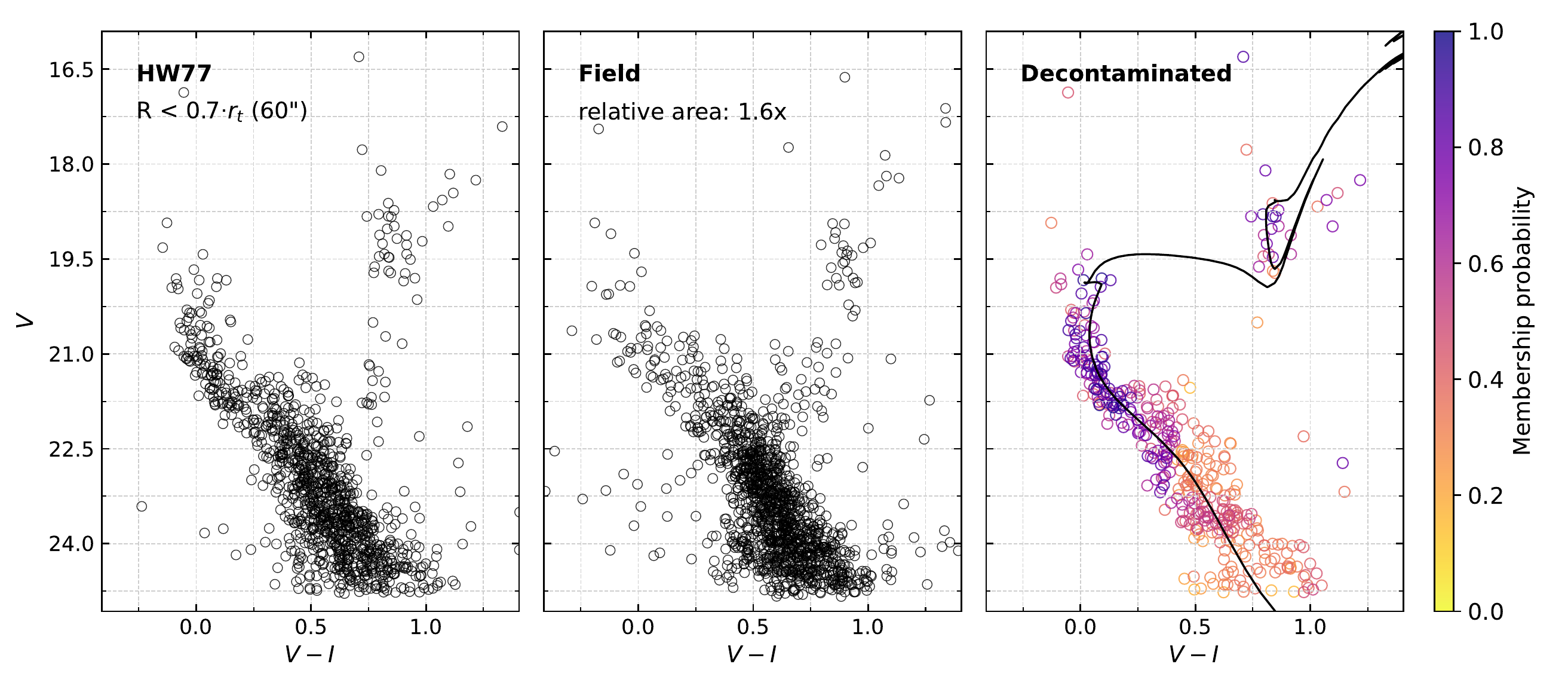}
    \caption{Example of the decontamination method for HW77. \textit{(Left:)}
    CMDs with the stars within $0.7 r_t$; \textit{(middle:)} field stars of a nearby sample ($r>72^{\prime\prime}$); \textit{(right:)} decontaminated CMD, with the best-fit isochrone of $1.12$\,Gyr and $\rm{[Fe/H]}=-1.02$ (see Table~\ref{tab:isoc-params}).}
    \label{fig:decont-hw77}
\end{figure*}

In Table~\ref{tab:literature}, we report the literature results on age and metallicity
for the sample clusters, including the compilation of \citet{2020AJ....159...82B} and subsequent work. The reported age determinations were mainly based on visual isochrone fits, and most of them fixed the metallicity and/or distance: \citet{2010A&A...517A..50G} and \citet{2014MNRAS.437.2005M} assumed $\rm{[Fe/H]}=-0.58$\,dex and $60.3$\,kpc; \citet{2007MNRAS.381L..84P, 2007MNRAS.382.1203P} assumed $\rm{[Fe/H]}=-0.7$\,dex and $56.8$\,kpc; and \citet{2011MNRAS.416L..89P, 2011MNRAS.418L..69P} and \citet{2011MNRAS.417.1559P, 2015MNRAS.450..552P} assumed $\rm{[Fe/H]}=-0.7$\,dex and $60.3$\,kpc.
The only papers which provided four free parameters were \citet{2015MNRAS.453.3190B}, \citet[][\texttt{ASteCA} code]{2017A&A...602A..89P} and \citetalias{2019MNRAS.484.5702M}, where the metallicities are derived from CMDs or CaT spectroscopy \citep{1998AJ....115.1934D, 2009AJ....138..517P, 2015AJ....149..154P, 2022A&A...664A.168D}. 
In this work, we present the isochrone fits to the observed data, indicating our results for age, metallicity, distance and reddening as free parameters, in some cases compared with available literature results.
\section{Methodology}
\label{sec3}

\subsection{Radial density profile: King models}

We have obtained the radial density profiles (RDPs) of the sample clusters by counting the number of stars in concentric rings around the centre. As a first step, the centre
of each cluster
is determined iteratively from the centroid of stellar coordinates
inside a 50 arcsec radius aperture,
starting with an initial guess and adjusting the new centre
and aperture
at each step.
Whenever convergence is achieved (i.e. centre coordinates change is inferior than 0.5 arcsec), the aperture size is reduced to 2/3 its current size and process is repeated, until aperture reaches $10-20$\,arcsec, depending on the cluster size and concentration. At this point, the RDP is checked to ensure that the maximum central density was reached.
The 4-parameter analytical model from \citet{1962AJ.....67..471K} was fitted to the observed profiles, according to the equation:
\begin{equation}
    \ \rho(r) = \rho_0 \left [ \frac{1}{\sqrt{1 + (r/r_c)^2}}  - \frac{1}{\sqrt{1 + (r_t/r_c)^2}} \right ]^2 + \rho_{bg} \; ,
\end{equation}
where $\rho_0$ is the projected central density, $r_c$ and $r_t$ are the core and tidal radii, and $\rho_{bg}$ is the projected density of background stars.

Different bin sizes were used to obtain the RDPs, with the smallest bin
size limited by
the cluster core radius (obtained iteratively).
The core and tidal radii, central and background density are obtained by
a $\chi^2$ minimisation method
to find the best-fitting model, comparing the observed and predicted density of stars
in each bin.
The best solution is usually uniquely defined and its uncertainties depends mostly on the star counts statistics and on the stellar background fluctuations.
This method was already applied to VISCACHA data in \citetalias{2019MNRAS.484.5702M} and \citet[\citetalias{2020MNRAS.498..205S}]{2020MNRAS.498..205S}.

In this work, since we are mainly interested in the cluster tidal radius, which gives us a measure of its size, we applied no correction for completeness. As discussed in \citetalias{2020MNRAS.498..205S}, the tidal radius
and cluster centre
derived from RDPs
are
largely unaffected by incompleteness, in contrast with the core radius. We will defer a more complete structural analysis of these clusters to a future work. Figure \ref{fig:rdp} shows the King profile fitting over the RDP of L113.

Although both the core and tidal radii are derived in the fitting procedure, we will only list the latter in Table \ref{tab:isoc-params}.
The fit did not converge for three sample clusters (HW59, B147 and HW82), for which the decontamination was carried out with a visual radius to limit the cluster sample.
As in \citetalias{2020MNRAS.498..205S}, we obtained a tidal radius larger than the SAMI field of view ($r_t\gtrsim 100$\,arcsec) for seven sample clusters,
that are probably biased toward smaller values, due to the reduced cluster coverage area (relative to the cluster full size) used in their derivation
\citep[see][]{2019A&A...625A.115O}.
For smaller clusters, it is possible to check that our RDP fitting have sufficient spatial coverage, by extending the sample to larger radii and verifying the stability of the derived tidal radii. 






\subsection{Statistical decontamination: photometric membership}

Since no
proper motion data
are available for the faint stars of the 
studied clusters, the membership probability was obtained from a statistical analysis as first described in \citet{2010MNRAS.407.1875M}.
The method compares probable cluster stars within the tidal radius with those in a nearby field. The distance of each star to the cluster centre and the local overdensity of stars in a CMD grid (compared with the same region in the field CMD) are used to define a cluster membership probability and clean the cluster sample from field stars. This method was also adopted in previous VISCACHA papers: \citetalias{2019MNRAS.484.5702M}, \citet[\citetalias{2021A&A...647L...9D}]{2021A&A...647L...9D}, \citet[\citetalias{2022MNRAS.512.4334D}]{2022MNRAS.512.4334D} and \citet{2022MNRAS.517L..41B}.

Figure~\ref{fig:decont-hw77} illustrates the method for the Bridge cluster HW77. In this case, the stars within a radius of $0.7 r_t$ were compared to a nearby field,
for which the lower main sequence (MS) and the red clump (RC) are similar to the cluster, but with an older main sequence turnoff (MSTO).
Therefore, the decontamination was very effective in removing field stars in $V\sim22$\,mag, which could deviate the isochrone fitting to an older MSTO. Also note the smooth luminosity function throughout the MS of the decontaminated sample.



\subsection{Statistical isochrone fitting}
\label{sec:3.3}

The \texttt{SIRIUS} code developed by our group \citep{2020ApJ...890...38S} applies a self-consistent Bayesian isochrone fitting to analyse the CMDs of stellar clusters. Through a Markov chain Monte Carlo (MCMC) algorithm, the code is able to derive for each cluster their probability distributions for age, metallicity, absolute distance modulus and reddening.
A likelihood function sums up a chi-square in magnitude and colour for each star compared to the closest isochrone point and divided by the photometric errors, as given below:
\begin{equation}
 \ \mathcal{L} = \sum^{N}_{i=1} \left [ - \chi^2_{\rm{mag},i} - \chi^2_{\rm{col},i} + \ln(p_{memb, i}) - \ln{(n_{*})} \right ] \; ,
 \label{eq:likel}
\end{equation}
where the index $i$ covers all the $N$ member stars, $mag$ and $col$ are the observed $V$ and $V-I$, $p_{memb}$ is the membership probability, and $n_*$ is the number of neighbouring stars in the CMD. The membership probability is implemented in such a way that higher-membership stars  contribute more in the fitting process. Similarly, the $n_*$ quantity is also considered to reduce the weight of stars in the more populated CMD regions, in particular the MS in favour of the more evolved sequences.

The Padova and TRieste Stellar Evolution Code set of isochrones \citep[PARSEC\footnote{\url{http://stev.oapd.inaf.it/cgi-bin/cmd_3.7}};][]{2012MNRAS.427..127B} was adopted, covering masses of $0.1$ to $350\,M_\odot$, metallicities of $-1.5 < \rm{[Fe/H]\:(dex)} < 0.0$ and ages between 1\,Myr and 10\,Gyr, which represent well the SMC stellar population. 
For clusters younger than $\sim 100$\,Myr, the pre-main sequence stage is clearly detected in the isochrones and CMDs (see Figure~\ref{fig:young-cmds}).
Our grid of isochrones has steps of $0.01$\,dex both in $\mathbf{\log(\rm{age})}$ and metallicity, but linear interpolations are carried out to get intermediate values.
A total to selective extinction ratio of $R_V=3.1$ 
is adopted.



The Python libraries \texttt{emcee} and \texttt{corner} \citep{2013PASP..125..306F, corner} were adapted to carry out the MCMC sampling and produce the corner plots with the posterior distributions in different projections.

Some prior distributions were employed depending on the cluster age. For clusters older than $\sim 1$\,Gyr with a well-defined
RC,
selected stars
in this stage
are used to match the
respective
isochrone points, thus helping constrain the distance and metallicity.
For the younger clusters, the isochrone MSTO, identified by the sharp reduction in the number of stars at the top of the MS, was used instead. In these cases, a compromise was obtained between an optimal fit of the MSTO and the entire MS, e.g. for L107 (Figure~\ref{fig:young-cmds}), HW82 and BS226.
For the very young clusters with few or no giants detected, the metallicity was first estimated from a
tentative fit without any prior information,
followed by a new fit with a gaussian prior,
in order
to obtain a better convergence and circumvent the degeneracy between $(m-M)_0$ and $\rm{[Fe/H]}$.



Following what was done in \citetalias{2021A&A...647L...9D} and \citetalias{2022MNRAS.512.4334D}, a metallicity prior was adopted for L91, L110, HW86 and L113, centred on the value derived from CaT spectroscopy given in Table~\ref{tab:literature}, in order to improve the age accuracy.
A limit of $\sim 1$\,Gyr is generally assumed as the minimum age for which the CaT is well calibrated \citep{2009AJ....138..517P}, which
is coherent with the limit between old clusters with red giants and young clusters, defined by \citet{1994AJ....107.1079P} and \citet{1994AJ....108.1773J} as those older than the Hyades \citep[787\,Myr;][]{1996ASPC...90..475M}.

Clusters parameters and uncertainties were derived by fitting a skewed gaussian over the  marginal posterior distribution of each parameter, as shown in Fig. \ref{fig:L114-results}.
Even though this method produces slightly larger uncertainties than those normally derived from 16th and 86th percentiles, we have found that it better matches the expected (i.e. peak) value of the parameters' distribution. The resulting values and uncertainties are shown in Table \ref{tab:isoc-params}.

\begin{table*}
	\centering
	\caption{Results of the new center and tidal radius obtained in the King profile fitting, fundamental parameters (age, metallicity, distance and reddening) obtained in the isochrone fitting with \texttt{SIRIUS}, and the integrated magnitude and total mass derived from the flux of member stars and a proper calibration. The values correspond to the median and $1\sigma$ level of the posterior distribution.
 }
	\label{tab:isoc-params}
	\begin{tabular}{lcccccccccc}
		\hline
        \multirow{2}{*}{Cluster} & RA$_{\rm{J2000}}$ & DEC$_{\rm{J2000}}$ & $r_t$ & Age & $\rm{[Fe/H]}$ & $d$ & $E(B-V)$ & $M_V$ & $\log(\mathrm{M}/M_\odot)$ \\ 
		& (h:m:s) & ($^\circ : ^\prime : ^{\prime\prime}$) & (arcsec) & (Gyr) & (dex) & (kpc) & (mag) & (mag) \\ 
		\hline

        HW55 & 01:07:20.0 & $-$73:22:39 & $113 \pm 31$ & $2.22 \pm 0.29$ & $-0.71\pm0.26$ & $64.3\pm5.0$ & $0.09\pm0.09$ & $-4.59\pm0.18$ & $3.58\pm0.13$ \\ 

        K55 & 01:07:32.5 & $-$73:07:14 & $99 \pm 16$ & $0.52\pm0.03$ & $-0.38\pm0.10$ & $52.0 \pm 1.7$ & $0.02 \pm 0.02$ & $-5.01\pm0.07$ & $3.37\pm0.10$ \\

        K57 & 01:08:14.0 & $-$73:15:27 &$73 \pm 10$ & $0.53\pm0.15$ & $-0.40\pm0.30$ & $53.0 \pm 3.4$ & $0.03 \pm 0.03$ & $-5.33\pm0.14$ & $3.50\pm0.14$ \\


        HW59 & 01:08:54.3 & $-$73:14:38 & $-$ & $6.8\pm2.4$ & $-0.99\pm0.35$ & $69.2\pm8.0$ & $0.09\pm0.08$ & $-4.01\pm0.26$ & $3.64\pm0.18$ \\ 

        HW63 & 01:10:12.3 & $-$73:12:32 & $103 \pm 25$ & $2.53\pm0.24$ & $-0.62\pm0.23$ & $67.6\pm6.5$ & $0.05\pm0.05$ & $-4.66\pm0.21$ & $3.64\pm0.14$ \\ 

        L92 & 01:12:33.0 & $-$73:27:24 & $58 \pm 6$ & $0.117 \pm 0.028$ & $-0.50\pm0.17$ & $54.7 \pm 4.8$ & $0.11 \pm 0.03$ & $-5.55\pm0.19$ & $3.19\pm0.14$ \\ 

        L93 & 01:12:48.2 & $-$-73:28:25 & $96 \pm 21$ & $3.02 \pm 0.30$ & $-0.70\pm0.22$ & $57.5 \pm 2.4$ & $0.06 \pm 0.04$ & $-4.63\pm0.09$ & $3.68\pm0.11$ \\ 

        L91 & 01:12:51.6 & $-$73:07:07 & $100 \pm 7$ & $3.9 \pm 0.5$ & $-0.82\pm0.08$ & $59.2\pm2.5$ & $0.15\pm0.03$ & $-5.39\pm0.09$ & $4.05\pm0.12$ \\ 

        B147
        & 01:14:50.5 & $-$73:06:49 & $-$ & $0.19\pm0.06$ & $-0.26\pm0.15$ & $50.0\pm4.1$ & $0.18\pm0.05$ & $-5.61\pm0.19$ & $3.33\pm0.15$ \\ 

        HW71se
        & 01:15:32.2 & $-$72:22:44 & $32 \pm 8$ & $0.16 \pm 0.05$ & $-0.53\pm0.21$ & $57.3 \pm 7.1$ & $0.03 \pm 0.05$ & $-5.07\pm0.27$ & $3.07\pm0.17$ \\ 

        HW75
        & 01:17:29.9 & $-$73:34:15 & $54 \pm 15$ & $0.11\pm0.03$ & $-0.56\pm0.13$ & $53.7\pm4.9$ & $0.13\pm0.05$ & $-5.08\pm0.21$ & $2.99\pm0.15$ \\ 

        HW77 & 01:20:11.0 & $-$72:37:19 & $86 \pm 15$ & $1.12 \pm 0.10$ & $-1.02\pm0.11$ & $58.3 \pm 3.2$ & $0.04 \pm 0.03$ & $-4.44\pm0.15$ & $3.35\pm0.12$ \\ 

        HW78
        & 01:21:20.7 & $-$73:05:40 & $123 \pm 36$ & $0.051\pm0.007$ & $-0.39\pm0.09$ & $53.0\pm3.9$ & $0.15\pm0.04$ & $-6.31\pm0.17$ & $3.27\pm0.12$ \\ 

        L101
        & 01:23:44.2 & $-$73:12:29 & $86 \pm 24$ & $0.013^{+0.025}_{-0.003}$ & $-0.27\pm0.13$ & $51.1\pm4.2$ & $0.09\pm0.04$ & $-5.74\pm0.18$ & $2.68\pm0.29$ \\ 
        
        HW81
        & 01:24:11.8 & $-$73:09:18 & $51 \pm 11$ & $0.004\pm0.001$ & $-0.11\pm0.18$ & $68.2\pm9.4$ & $0.21\pm0.04$ & $-7.53\pm0.30$ & $3.09\pm0.16$ \\ 

        HW82
        & 01:24:27.7 & $-$73:10:16 & $-$ & $0.050\pm0.013$ & $-0.41\pm0.15$ & $61.4\pm7.6$ & $0.13\pm0.05$ & $-6.69\pm0.28$ & $3.42\pm0.16$ \\ 

        L104
        & 01:25:26.1 & $-$73:23:17 & $68 \pm 27$ & $0.030\pm0.007$ & $-0.25\pm0.16$ & $65.8\pm6.1$ & $0.08\pm0.04$ & $-6.95\pm0.20$ & $3.39\pm0.14$ \\ 

       B165 & 01:30:50.5 & $-$73:26:03 & $55 \pm 41$ & $0.33\pm0.08$ & $-0.54\pm 0.18$ & $52.0\pm7.4$ & $0.01\pm0.05$ & $-3.86\pm0.31$ & $2.79\pm0.17$ \\ 

        BS187 & 01:31:01.8 & $-$72:51:01 & $67 \pm 26$ & $1.01 \pm 0.22$ & $-0.92\pm0.15$ & $52.7 \pm 3.9$ & $0.15 \pm 0.05$ & $-4.36\pm0.17$ & $3.28\pm0.13$ \\ 

        L107
        & 01:31:06.7 & $-$73:24:45 & $75 \pm 10$ & $0.013 \pm 0.005$ & $-0.41\pm0.17$ & $55.7\pm7.7$ & $0.06\pm0.04$ & $-7.03\pm0.31$ & $3.20\pm0.18$ \\ 

        L109 & 01:33:14.3 & $-$74:09:58 & $78 \pm 9$ & $4.06 \pm 0.52$ & $-0.79\pm0.21$ & $58.6 \pm 2.7$ & $0.09 \pm 0.04$ & $-4.15\pm0.10$ & $3.56\pm0.12$ \\ 

        L110 & 01:34:26.0 & $-$72:52:28 & $101 \pm 3$ & $5.0\pm0.7$ & $-0.94\pm0.11$ & $61.7\pm4.3$ & $0.07\pm0.04$ & $-5.59\pm0.15$ & $4.19\pm0.13$ \\ 

        HW86 & 01:42:23.3 & $-$74:10:28 & $96 \pm 15$ & $1.46 \pm 0.10$ & $-0.69\pm0.12$ & $51.3\pm2.8$ & $0.08\pm0.04$ & $-4.13\pm0.13$ & $3.29\pm0.12$ \\ 

        WG1
        & 01:42:52.7 & $-$73:20:09 & $132 \pm 18$ & $0.031\pm0.005$ & $-0.23\pm0.14$ & $54.5\pm4.8$ & $0.20\pm0.04$ & $-5.94\pm0.20$ & $2.99\pm0.13$ \\ 

        BS198
        & 01:47:57.9 & $-$73:07:47 & $98 \pm 34$ & $0.011 \pm 0.005$ & $-0.37\pm0.22$ & $58.3 \pm 7.5$ & $0.15 \pm 0.06$ & $-3.97\pm0.29$ & $1.93\pm0.19$ \\ 

        L113 & 01:49:30.3 & -73:43:40 & $132 \pm 5$ & $3.9 \pm0.4$ & $-0.87\pm0.08$ & $54.7\pm1.3$ & $0.03\pm 0.03$ & $-5.62\pm0.05$ & $4.14\pm0.11$ \\ 

        L114 & 01:50:19.3 & $-$74:21:21 & $87 \pm 4$ &  $0.033 \pm 0.004$ & $-0.47\pm0.08$ & $54.7 \pm 3.5$ & $0.03 \pm 0.03$ & $-6.77\pm0.15$ & $3.34\pm0.11$ \\ 

        NGC\,796
        & 01:56:44.6 & $-$74:13:10 & $78 \pm 6$ & $0.036 \pm 0.003$ & $-0.22\pm0.06$ & $60.5 \pm 3.6$ & $0.03 \pm 0.02$ & $-6.58\pm0.13$ & $3.29\pm0.11$ \\ 

        WG13
        & 02:02:40.9 & $-$73:56:23 & $53 \pm 11$ & $0.33\pm0.11$ & $-0.28\pm0.15$ & $55.0\pm7.3$ & $0.07\pm0.06$ & $-5.78\pm0.31$ & $3.55\pm0.18$ \\ 

        BS226 & 02:05:41.9 & $-$74:22:53 & $35 \pm 11$ & $1.09 \pm 0.21$ & $-1.12\pm0.17$ & $53.5\pm4.9$ & $0.07\pm0.05$ & $-3.50\pm0.21$ & $2.95\pm0.14$ \\ 

        ICA45
        & 02:27:13.3 & $-$73:45:27 & $68 \pm 6$ & $0.021 \pm 0.005$ & $-0.38\pm0.15$ & $58.3\pm5.6$ & $0.08\pm0.04$ & $-5.48\pm0.21$ & $2.70\pm0.14$ \\ 

        BS245 & 02:27:27.6 & $-$73:58:27 & $38 \pm 2$ & $0.10\pm0.03$ & $-0.55\pm0.25$ & $50.6\pm7.9$ & $0.10\pm0.07$ & $-5.36\pm0.36$ & $3.08\pm0.18$ \\ 

        OGLB33 & 02:41:03.6 & $-$73:15:12 & $74 \pm 8$ & $4.0\pm1.1$ & $-0.75\pm0.17$ & $66.1\pm6.1$ & $0.12\pm0.07$ & $-2.24\pm0.23$ & $2.80\pm0.16$ \\ 

    \hline
		
	\end{tabular}
\end{table*}

\subsection{Integrated magnitudes and masses}

To derive the total masses for the clusters, we determined their integrated apparent $V$ magnitudes ($V_{int}$) by adding up the member star fluxes.  We then converted $V_{int}$ to the absolute one ($M_V$) by using the clusters' individual distance and extinction. Finally, the mass and its uncertainty was calculated following the calibration with age and metallicity
(fixed at $\rm{[Fe/H]}=-0.58$\,dex)
of simple stellar population models given by \citet{2014MNRAS.437.2005M}. Mass uncertainty comes from propagation of errors in $M_V$, $\mathbf{\log(\rm{age})}$, extinction and distance. The derived integrated magnitudes and masses for the clusters are shown in Table \ref{tab:isoc-params}.

\section{Results}
\label{sec4}

The isochrone fitting of the sample cluster
CMDs
provide
precise age, metallicity, distance and reddening. We also report revised coordinates and tidal radius based on the
RDPs.
These results are given in Table~\ref{tab:isoc-params}, together with the uncertainties.



\begin{figure*}
    \centering
    \includegraphics[trim={0 0.48cm 2.55cm 0},clip, height=8.13cm]{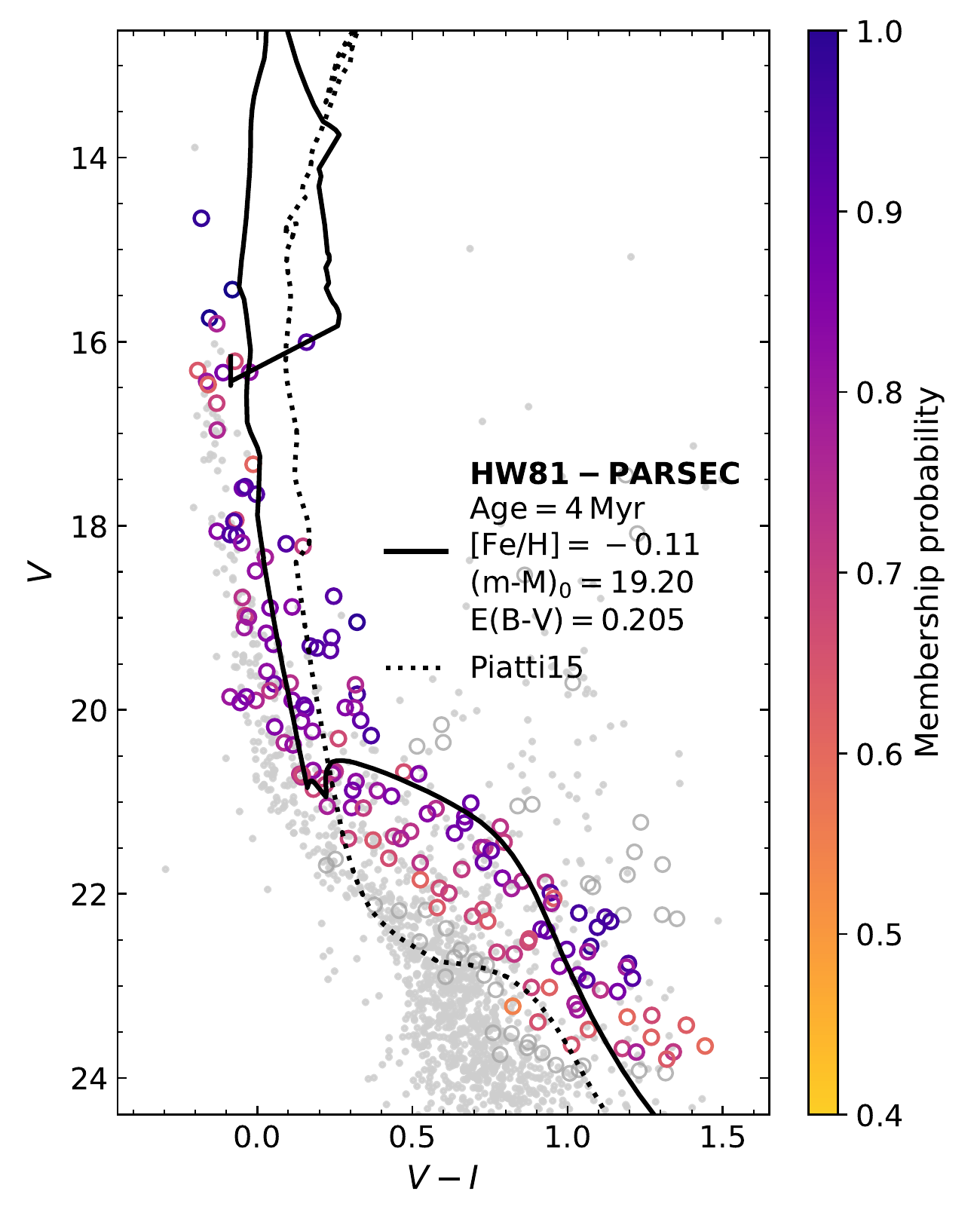}
    \includegraphics[trim={0 0.48cm 2.55cm 0},clip, height=8.13cm]{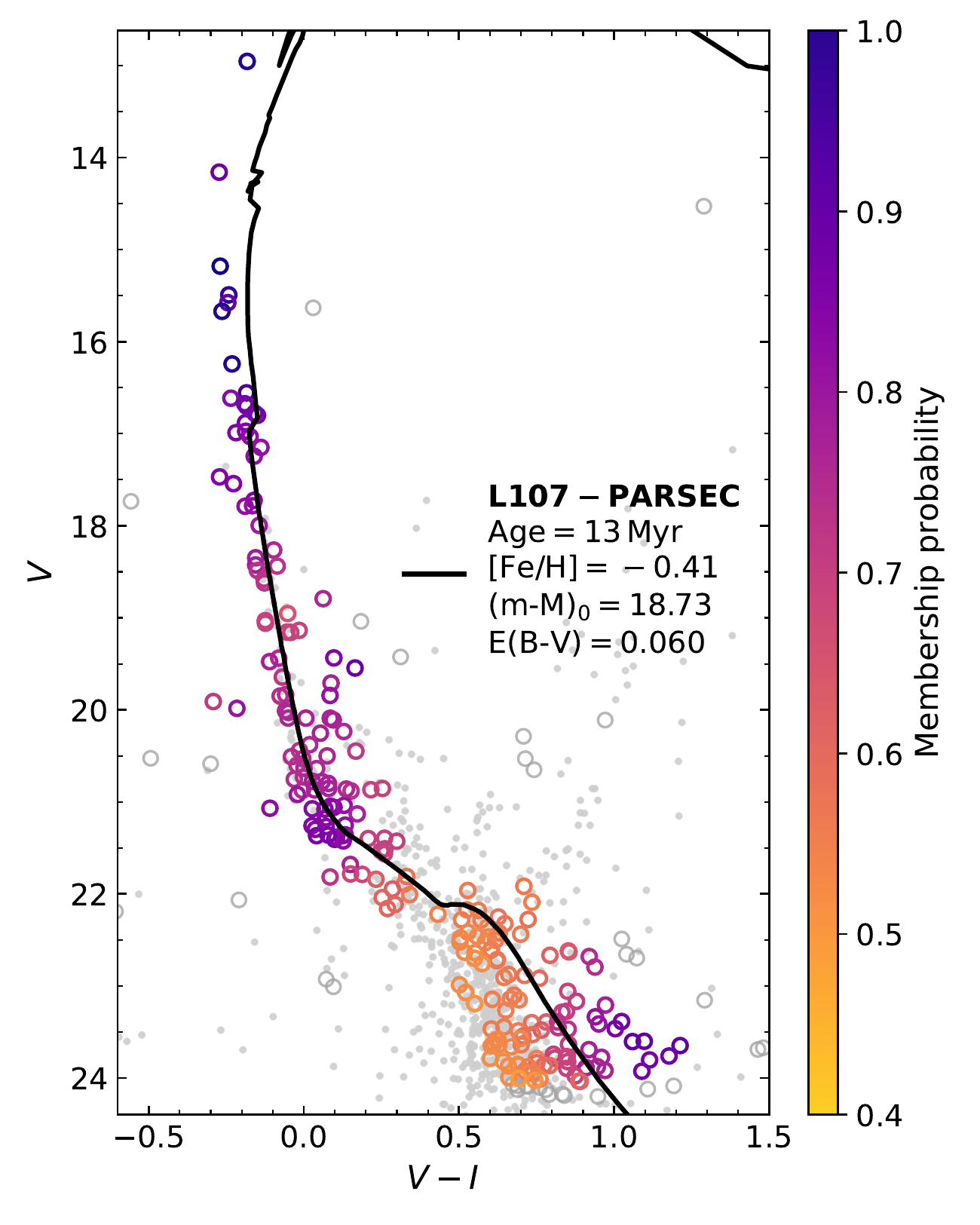}
    \includegraphics[trim={0 0.48cm 0 0},clip, height=8.13cm]{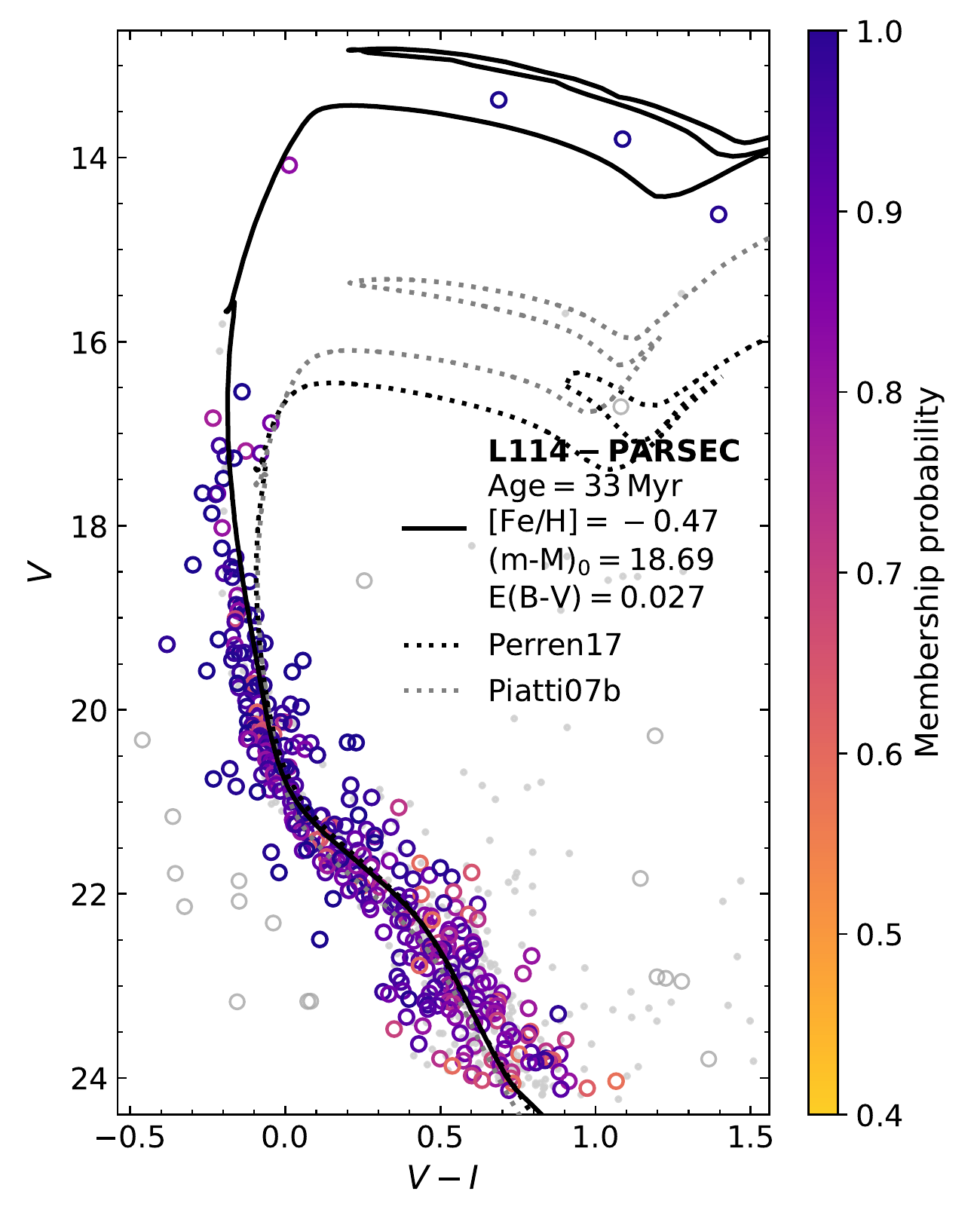}
    \caption{$V$ vs. $V-I$ decontaminated CMD of three of the youngest sample clusters, with the best-fit isochrone and a comparison with literature results.}
    \label{fig:young-cmds}
\end{figure*}

\begin{figure*} 
    \centering
    \includegraphics[height=8.6cm]{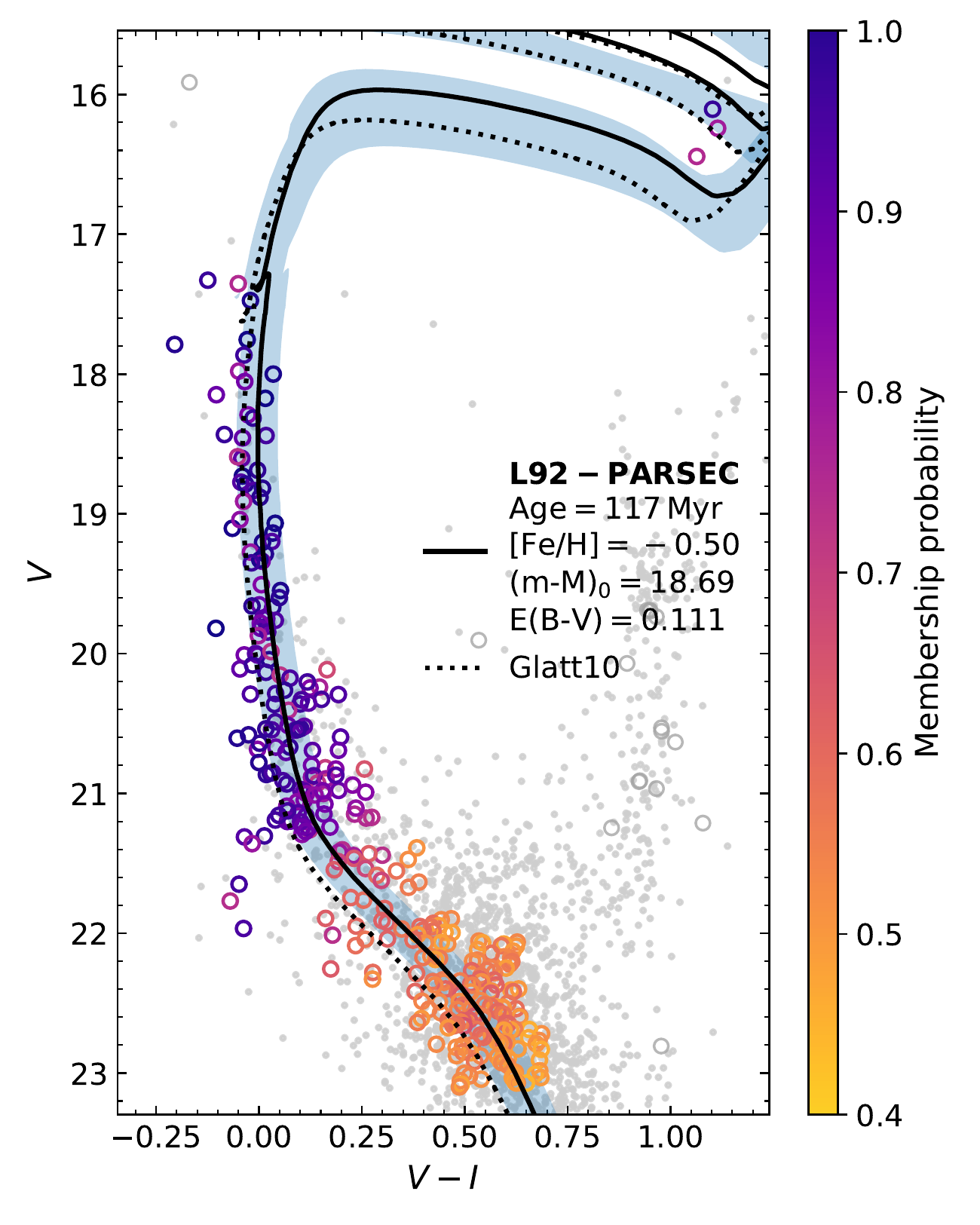}
    \hspace{2mm}
    \includegraphics[height=8.6cm]{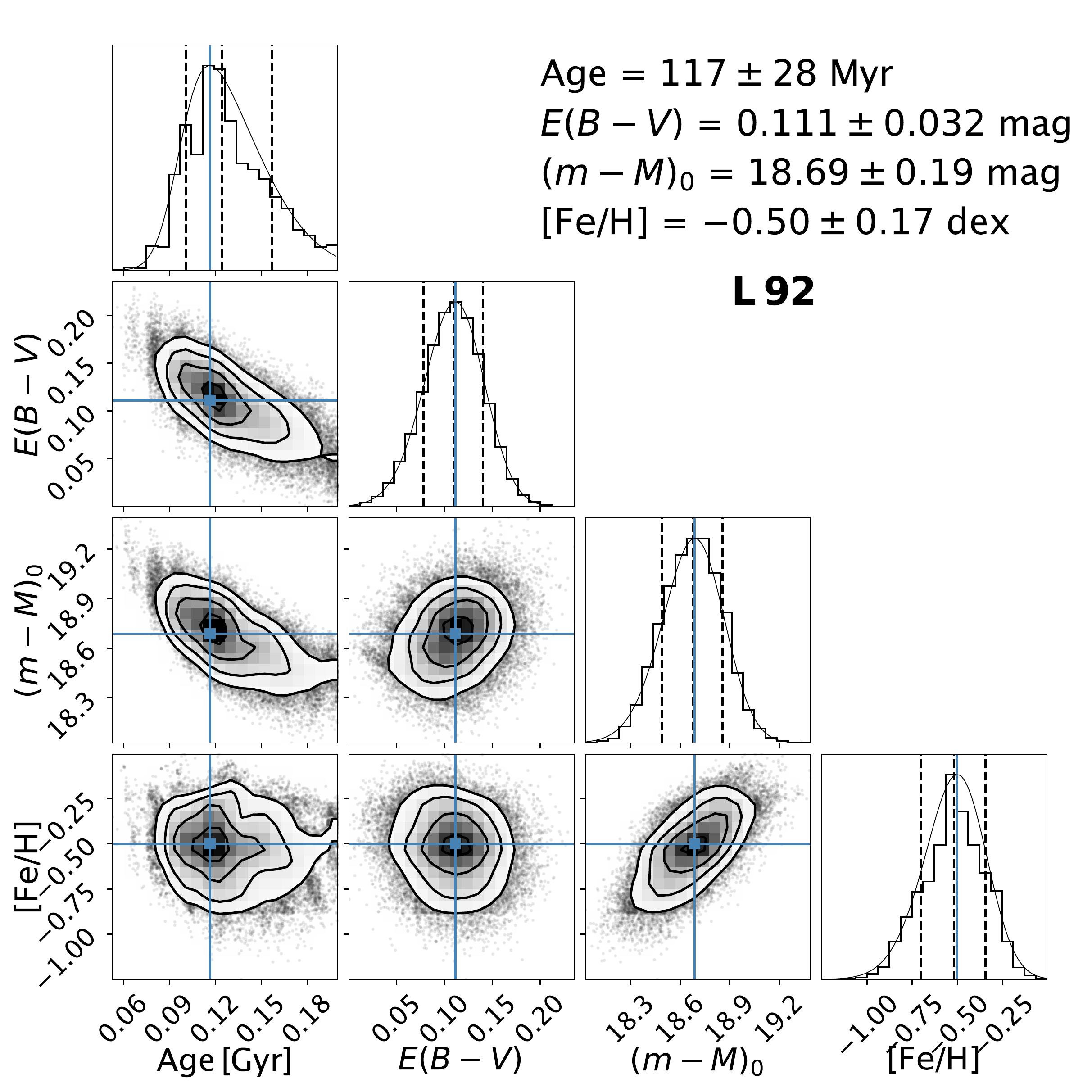}
    \caption{\textit{(Left:)} $V$ vs. $V-I$ decontaminated CMD of the young cluster L92 with the best-fit isochrone and a literature comparison \citep{2010A&A...517A..50G},
    as well as a shaded blue area containing the solutions within $1\sigma$.
    \textit{(Right:)} Corresponding corner plots, showing the posterior distribution and correlations between the parameters (dashed lines give the 16th, 50th and 86th percentiles, and the blue line marks the centre of the fitted skewed gaussian as the solution).}
    \label{fig:L114-results}
\end{figure*}

Figure~\ref{fig:young-cmds} shows the isochrone fits for three of the youngest sample clusters (HW81, L107 and L114), illustrating the vertical MS and the lack of giant stars, whereas the pre-main sequence is clearly sampled.
In these cases, the pre-main sequence turn-on provides the strongest constraint to the fitting procedure. Nevertheless, the relative uncertainties of the derived parameters for these clusters are higher than the sample average;
these uncertainties
should be considered as internal errors
(see Section~\ref{sec:3.3}).
Figure~\ref{fig:L114-results} shows the young cluster L92, which, differently than HW81 and L107, contains some giant stars that help constraining the metallicity. In this figure, we also show the resulting corner plot, which contains the posterior distribution in each parameter in the diagonal panels and the correlation between them in the other panels, as indicated by the labels.
In Figure~\ref{fig:old-cmds}, the isochrone fits of three old clusters are given, with a well-defined MS reaching $V\sim24$\,mag,
and well-populated giant branches and red clump, that helped to constrain the metallicity and distance.
The isochrone fits for the remaining clusters are presented in Appendix~\ref{app:CMDs}.

\subsection{Clusters with first derivation of age and metallicity}

This work contains the first age derivation
for nine sample clusters, namely: HW78, L101,
L104, B165, L107, WG1, BS198, ICA45 and OGLB33. HW78 was analysed by \citet{2015MNRAS.450..552P} with near-infrared
data, but was marked as a possible non-cluster.
Also note that L107 was presented in the \citet{1981A&AS...46...79V} compilation of integrated $UBV$ photometry with an age estimation of $<4$\,Myr (as well as $<200$\,Myr for L114 and $>1$\,Gyr for L113). OGLB33 was only recently catalogued by \citet{2017AcA....67..363S} as OGLE-MBR-CL-0033, without any further analysis. All the above mentioned clusters
contain very few member stars, where eight of them are very young ($<50$\,Myr old) and the other two are older and compact (B165 is 330\,Myr old and OGLB33 is 4.0\,Gyr old).

For these clusters and another nine of them (see Table~\ref{tab:literature}), totalling eighteen clusters,
metallicity was derived for the first time.
For the remaining fifteen clusters, the metallicity values available in the literature are from CaT spectroscopy for four clusters \citep{2009AJ....138..517P, 2015AJ....149..154P, 2022A&A...664A.168D, 1998AJ....115.1934D} and from CMD analysis for eleven of them \citep[and \citetalias{2019MNRAS.484.5702M}]{2015MNRAS.453.3190B, 2017A&A...602A..89P}.

A comparison between the derived metallicities and the mean metallicity from \citet{2020AJ....159...82B} is presented in Figure~\ref{fig:comp-lit}, showing a good agreement within the uncertainties. The largest discrepancy is observed for L114 \citep[$\rm{[Fe/H]}=-0.47$\,dex, compared to $-0.10$\,dex from][]{2017A&A...602A..89P}, followed by NGC796, K55 and L113.



\subsection{Comments on specific clusters}

In this section, we provide further comments on clusters that have conflicting results with the literature or particularities detected in the images and/or in the CMD analysis. 

\subsubsection*{HW59}

The old, compact cluster HW59 contains several blue, young stars in the surrounding field. The radial density profile was very noisy and no King function was fitted to the data, and a very small radius of $17^{\prime\prime}$ was adopted in the decontamination procedure to define the initial cluster sample, in order to avoid the presence of field stars. An age of $6.8\pm2.4$\,Gyr and a distance of $\sim 69\pm8$\,kpc were derived, making it the oldest and farthest cluster in our sample.


\subsubsection*{HW63 and L91}

\begin{figure*}
    \centering
    \includegraphics[trim={0 0.48cm 2.55cm 0},clip, height=8.13cm]{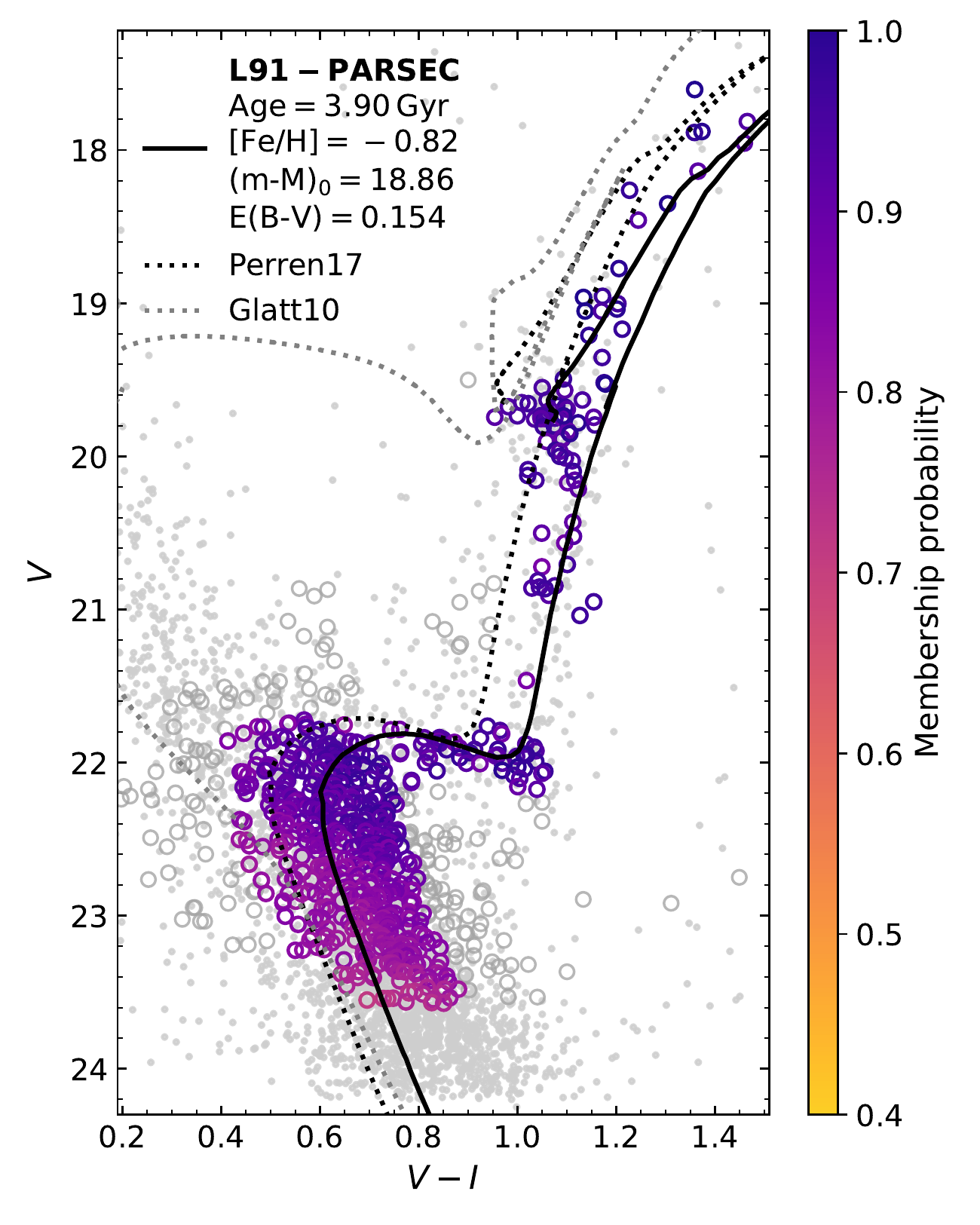}
    \includegraphics[trim={0 0.48cm 2.55cm 0},clip, height=8.13cm]{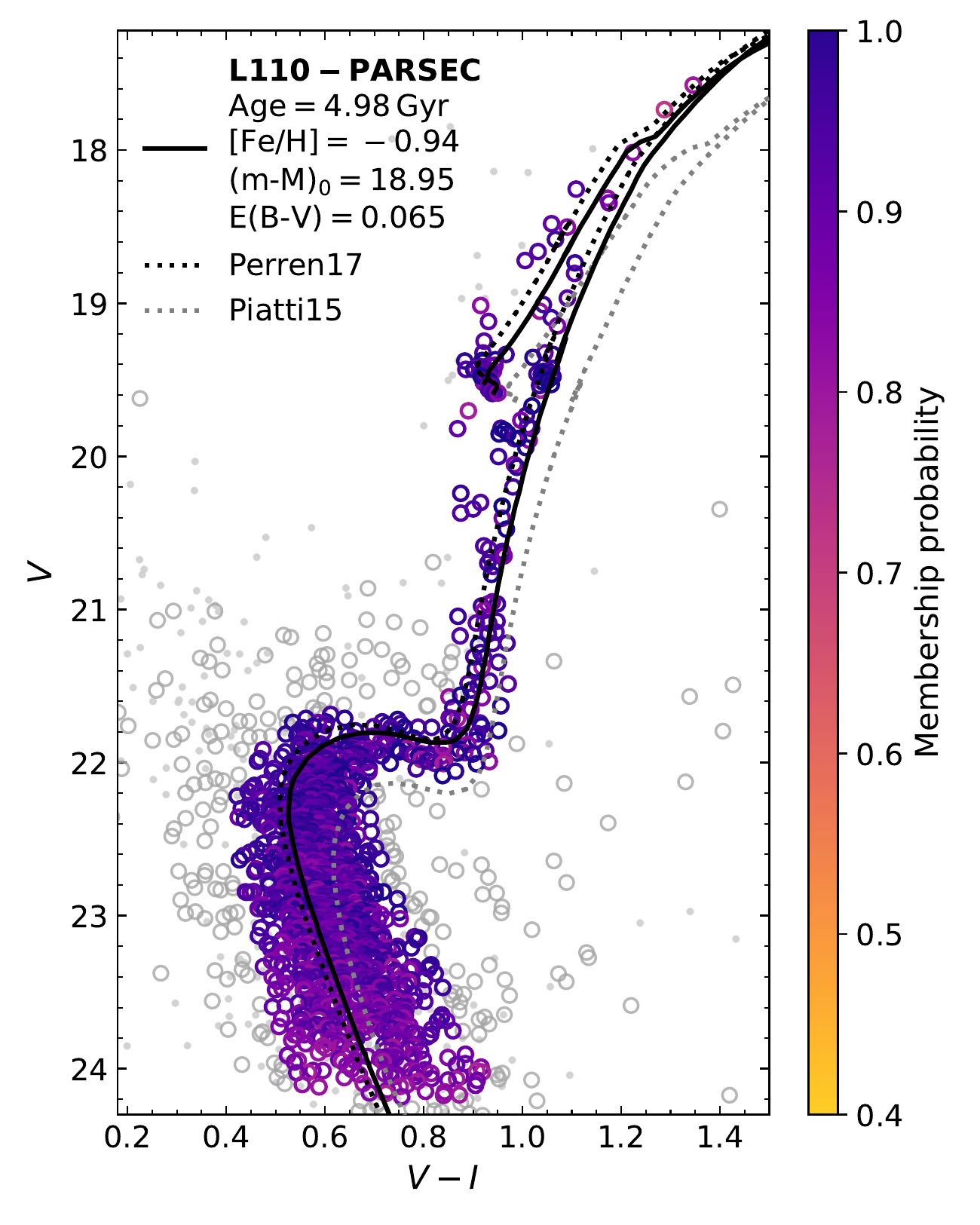}
    \includegraphics[trim={0 0.48cm 0 0},clip, height=8.13cm]{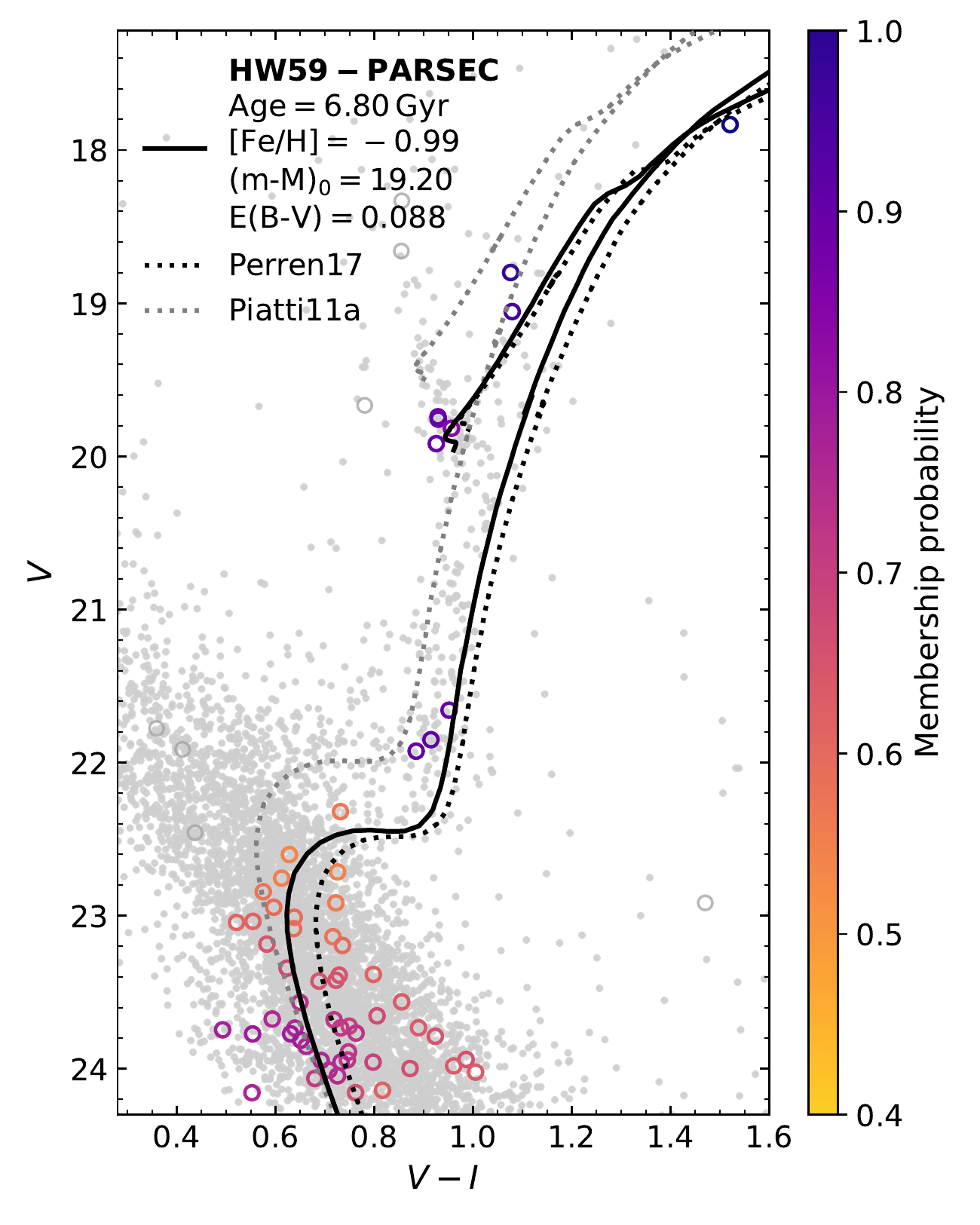}
    \caption{$V$ vs. $V-I$ decontaminated CMD of three old sample clusters, with the best-fit isochrone and a comparison with literature results.}
    \label{fig:old-cmds}
\end{figure*}

\begin{figure}
    \centering
    \includegraphics[width=0.46\textwidth]{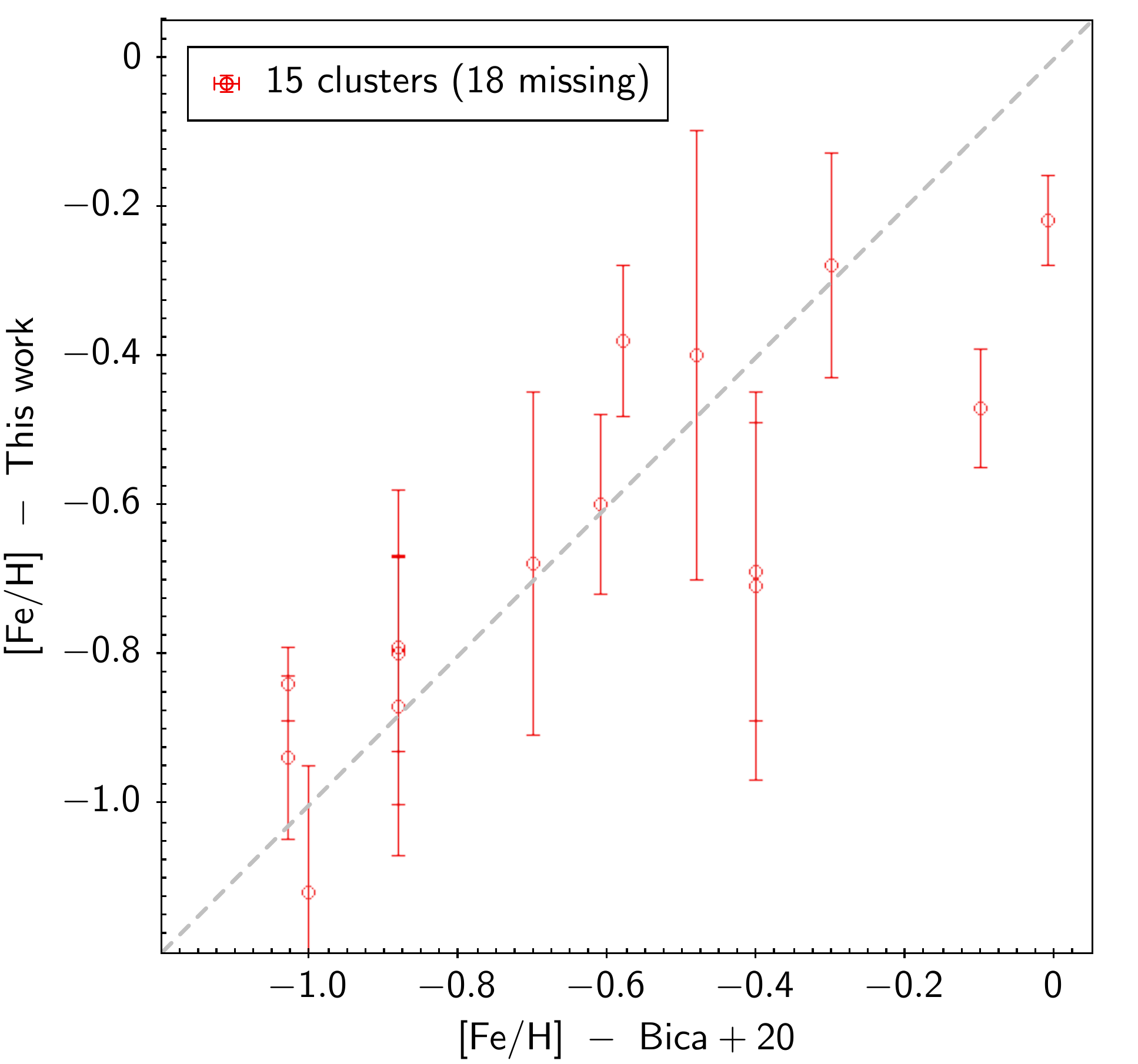}
    \caption{Comparison between the derived metallicities and the metallicity given in \citet{2020AJ....159...82B}, which consists of an average between literature values
    for 15 clusters
    (Table~\ref{tab:literature}).
    }
    \label{fig:comp-lit}
\end{figure}

There is a large difference in the literature ages for these clusters: \citet{2010A&A...517A..50G} derived young ages of $0.45\pm0.31$ and $0.79\pm0.55$\,Gyr for HW63 and L91 respectively, with MCPS photometry ($V\lesssim21$\,mag).
From $CT_1$ Washington photometry for both
($V \lesssim 22.5$\,mag), \citet{2011MNRAS.416L..89P} found $5.4\pm1.0$ 
and $4.3\pm1.0$\,Gyr, and \citet[\texttt{ASteCA} tool]{2017A&A...602A..89P} 
found $3.5\pm0.5$\,Gyr and $4.0\pm0.6$\,Gyr, for HW63 and L91 respectively.
In fact, \citet{2010A&A...517A..50G} states that their photometric limit hampered the age derivation for clusters older than 1\,Gyr, because the MSTO was not resolved. Our results confirm old ages of $2.5\pm0.2$ 
and $3.9\pm0.5$\,Gyr (Figure~\ref{fig:old-cmds}) for HW63 and L91, more compatible with \citet{2017A&A...602A..89P} and \citet{ 2011MNRAS.416L..89P}. Note that both clusters present a MSTO at $V\sim22$\,mag, requiring deep observations. 


\subsubsection*{HW77}
Figure~\ref{fig:decont-hw77} shows the deepest and best-seeing CMD of the sample, reaching $V\sim24.5$\,mag and $\rm{FWHM}=0.3$\,arcsec. With a derived metallicity of $\rm{[Fe/H]}=-1.02\pm0.11$\,dex, HW77 is among the most metal-poor sample clusters,
very close to the average SMC metallicity of the old field stellar population \citep{2018MNRAS.475.4279C},
and located in the Wing.

\begin{figure*}
    \centering
    \includegraphics[width=0.92\textwidth]{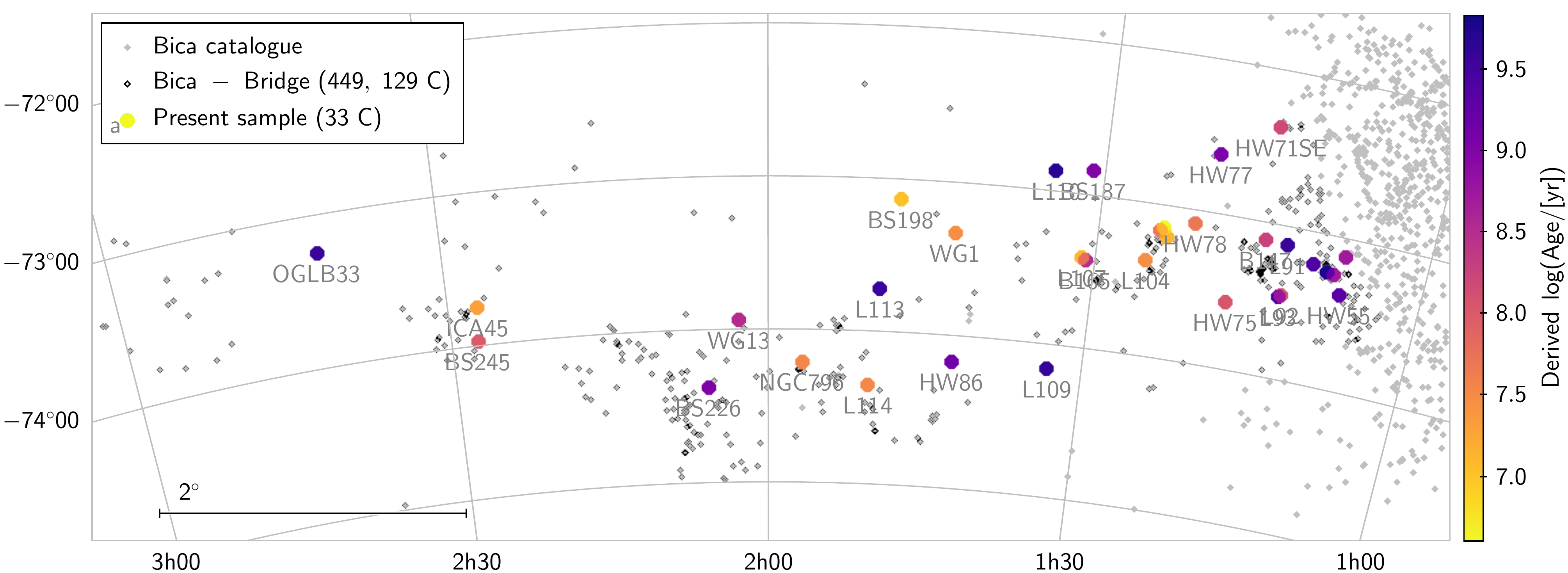}
    \includegraphics[width=0.95\textwidth]{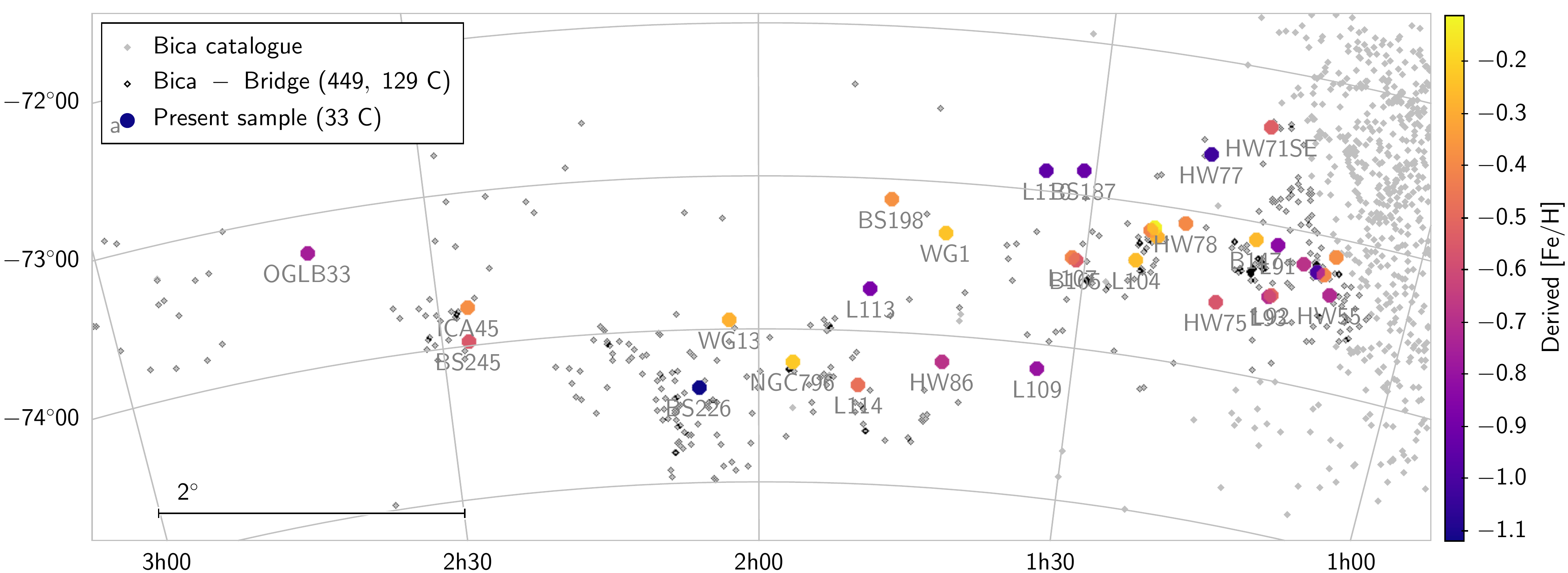}
    \caption{Projected distribution of the 33 sample clusters,
    colour-coded by the derived age (top panel) and $\rm{[Fe/H]}$ (bottom panel). The points are overplotted on the \citet{2020AJ....159...82B} catalogue (grey dots) and the 449 Bridge objects (black diamonds, of which 129 are classified as clusters).
    }
    \label{fig:spatial-ages}
\end{figure*}

\begin{figure}
    \centering
    \includegraphics[width=0.98\columnwidth]{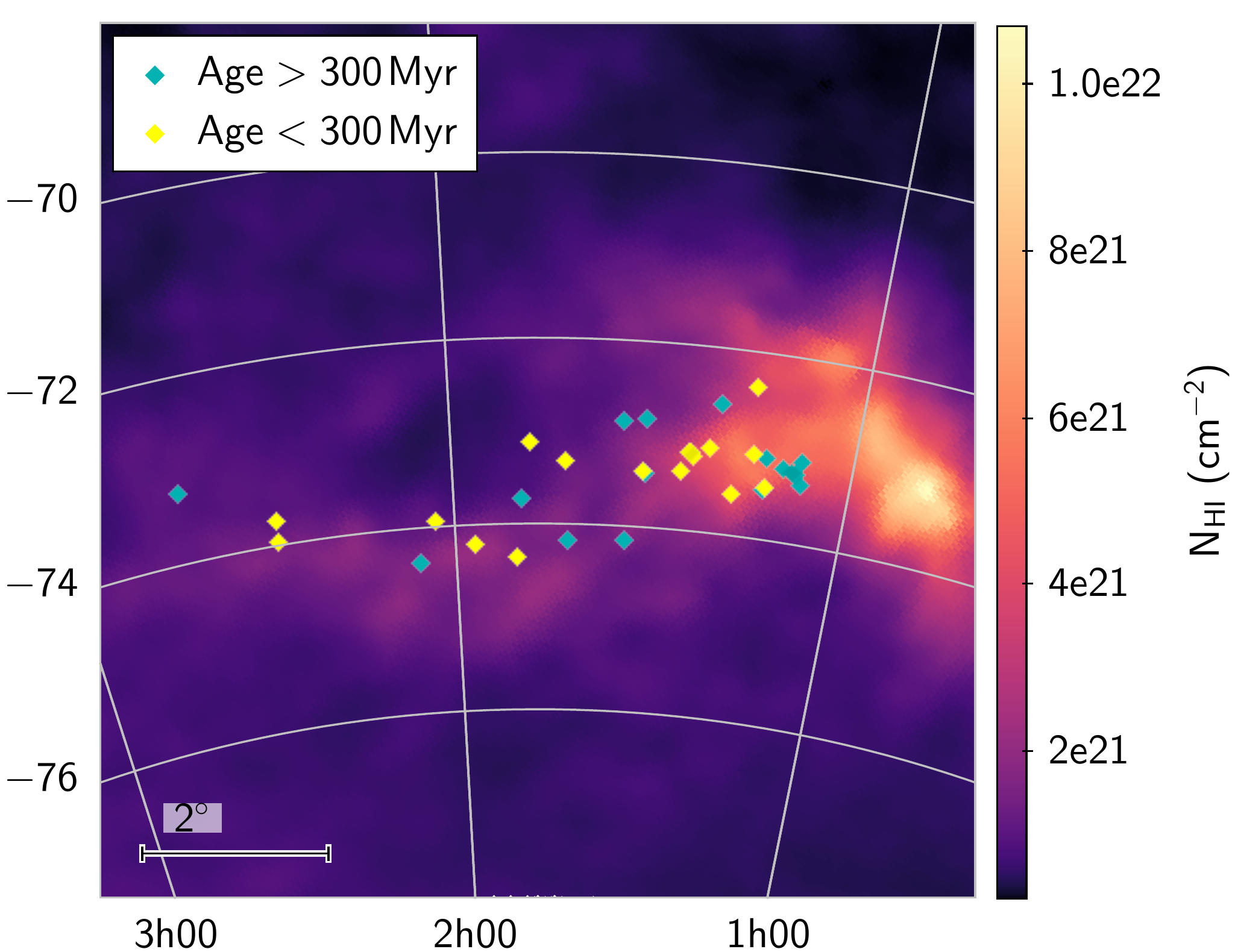}
    \caption{Map of HI column density from the HI4PI survey \citep{2016A&A...594A.116H}, with the Wing/Bridge and sample clusters overplotted. North is up and East to the left.} 
    \label{fig:hi-map-ages}
\end{figure}

\subsubsection*{HW81}
HW81 is particularly young, resulting in $4\pm 1$\,Myr and a relatively high reddening of $E(B-V)=0.21\pm0.04$\,mag. The CMD contains several pre-main sequence stars up to $V\sim 21$\,mag (Figure~\ref{fig:young-cmds}) and the composite image shows a blurry green region (Figure~\ref{fig:RGB_L113}), possibly indicating presence of some gas and dust of its embedded phase \citep[$< 5$\,Myr;][]{2018MNRAS.476..842O}.
This confirms the findings by \citet{2015MNRAS.450..552P} that HW81 is embedded in the HII region N88, as previously investigated using the adaptive optics system of the Very Large Telescope \citep{2010A&A...510A..95T}.

\subsubsection*{L113}
\citet{2021A&A...647A.135N} used Str\"omgren photometry
from the SOAR optical
imager
(SOI), obtaining $\rm{[Fe/H]}=-1.14 \pm 0.10$\,dex and an age in the range $3.5$ to 4 Gyr for this cluster.
\citet{2015AJ....149..154P} derived $\rm{[Fe/H]}=-1.03 \pm 0.04$\,dex from CaT lines. \citet{2010A&A...520A..85D} used integrated spectra to derive metallicity in the range $-2.1 < \rm{[M/H]\:(dex)} < -1.4$ and old ages. A metallicity of $\sim -1.0$\,dex as obtained here, as well as by \citet{2021A&A...647A.135N} and \citet{2015AJ....149..154P}, is more compatible with the SMC metallicity, therefore we discard the possibility of having a more metal-poor value.


\subsubsection*{L114}
The isochrones of 140 and 160\,Myr for this cluster found by \citet{2007MNRAS.382.1203P} and \citet{2017A&A...602A..89P} respectively, are plotted in Figure~\ref{fig:young-cmds} together with the one of $33$\,Myr found here. This figure shows that, in our decontaminated CMD, the literature values do not fit the brighter member giants.
Curiously, our CMD shows $3$ fainter giant stars near $V\sim16$\,mag, that might have biased the previous results to older ages. In our analysis they have been marked as field giants; furthermore the upper MS shape and bright (super)giants also appears to corroborate our results.

\subsubsection*{NGC796}
We obtained a distance of $60.5\pm3.6$\,kpc in very good agreement with \citet[$58.9\pm0.8$\,kpc]{2018ApJ...857..132K} and \citetalias{2019MNRAS.484.5702M} ($60.3^{+2.7}_{-2.4}$\,kpc).
The latter was obtained with the same VISCACHA data, but using a different method based on synthetic isochrones. This confirms that the distance from \citet{2015MNRAS.453.3190B} was too small, which may have resulted from shallower and noisier data from the SOI@SOAR imager without adaptive optics. 


\subsubsection*{BS226}
This cluster was previously analysed by \citet{2015MNRAS.453.3190B}, who found similar age and metallicity (see Table~\ref{tab:literature}), but like in the case of NGC796, a smaller distance of $\sim 40$\,kpc was derived. The present results are more reliable
as they are based on deeper photometry and adaptive optics. Our CMD presents
well-defined sequences showing that the lower MS and the few giants are better fitted with a larger distance of $\sim53$\,kpc.



\begin{figure*} 
    \centering
    \includegraphics[trim={2.55cm 1.85cm 2.55cm 0.45cm},clip, height=4.95cm]{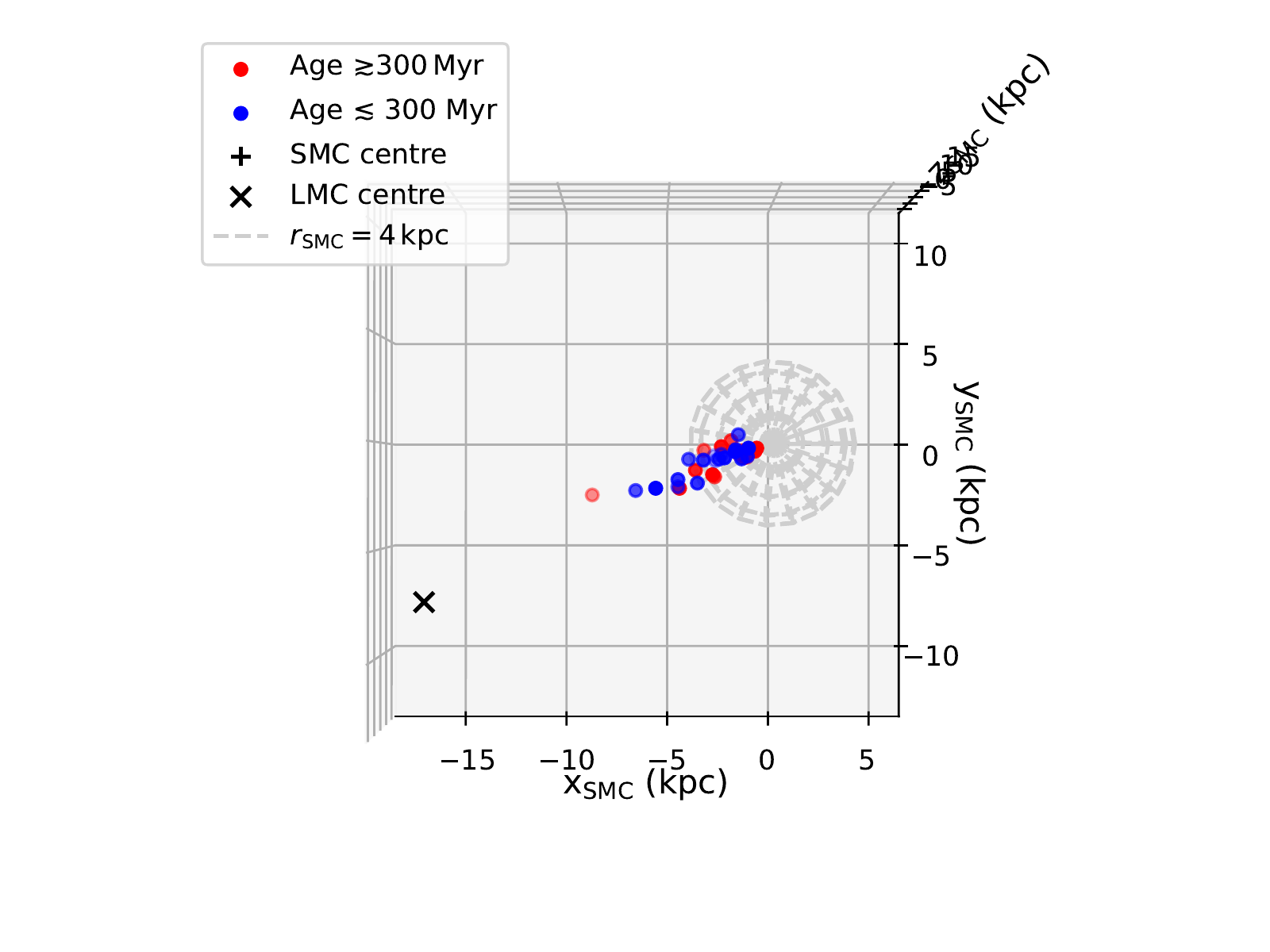}
    \includegraphics[trim={2.15cm 0.3cm 3.1cm 1.3cm},clip, height=4.95cm]{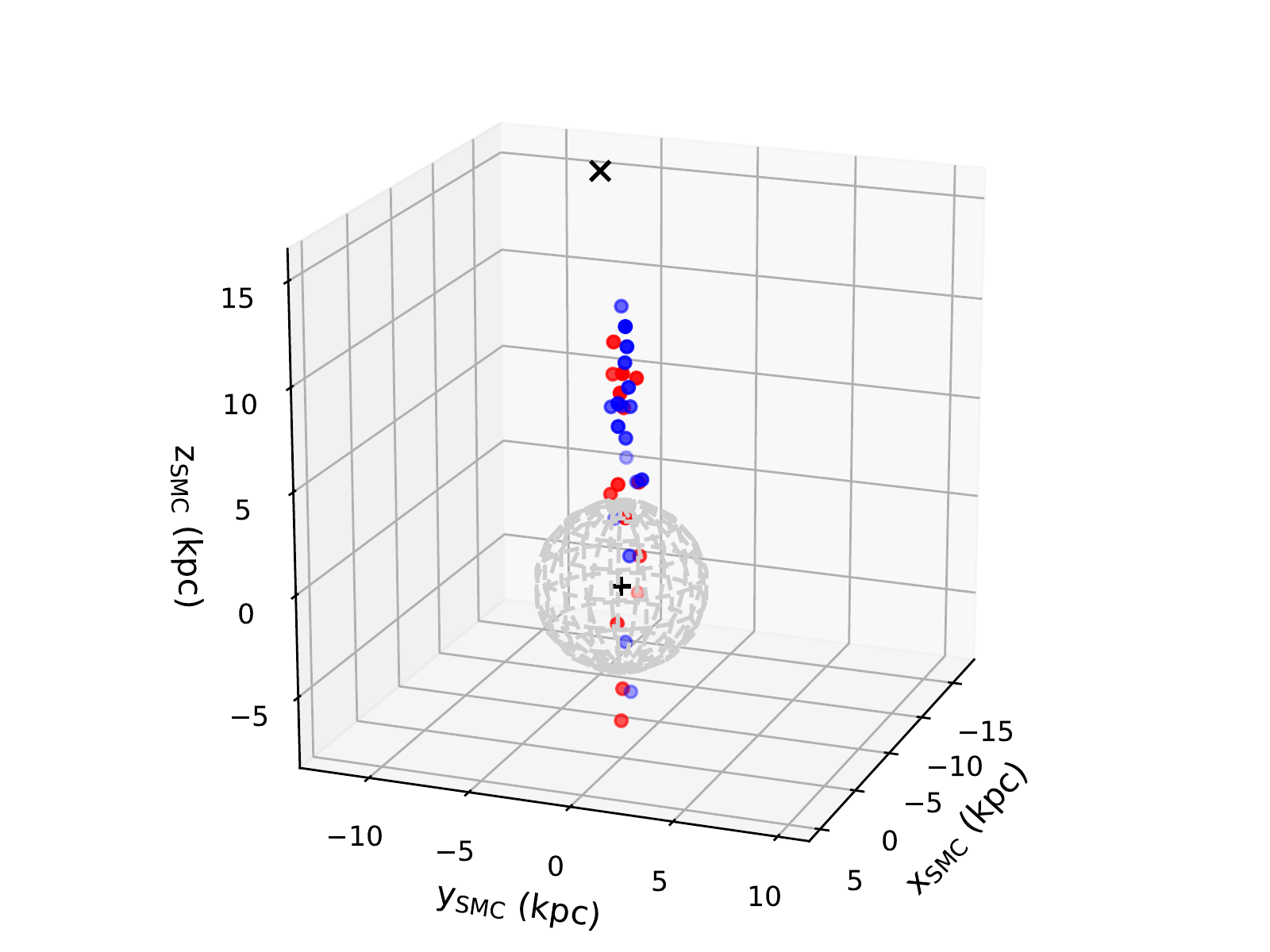}
    \includegraphics[trim={3.8cm 0.8cm 3.1cm 2.4cm},clip, height=4.95cm]{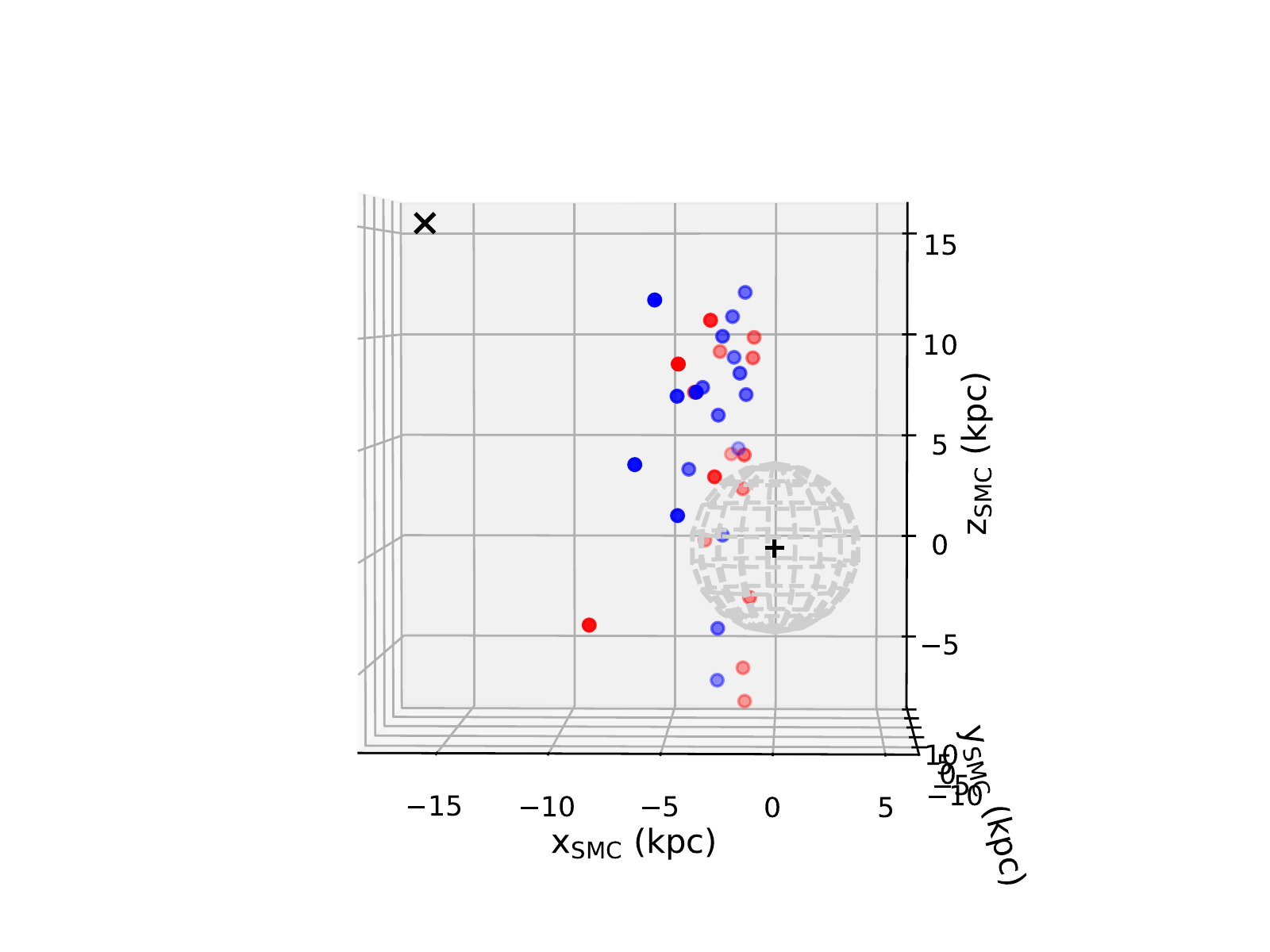}
    \caption{Projections of the three-dimensional distribution of the 33 sample clusters, identifying the old and the young clusters as red and blue symbols: \textit{(left:)} $x$ vs. $y$ similar to the sky projection; \textit{(middle:)} projection showing the alignment between SMC, LMC and Bridge clusters with different depth; \textit{(right:)} $x$ vs. $z$ projection. 
    The SMC is located at the origin, the LMC is at $(x,y,z) = (-16.0, -7.2, 15.2)$, and the SMC tidal radius of 4\,kpc is also shown.}
    \label{fig:mm0vsRA}
\end{figure*}

\begin{figure}
    \centering
    \includegraphics[trim={0cm 0.15cm 0cm 0},clip,width=0.98\columnwidth]{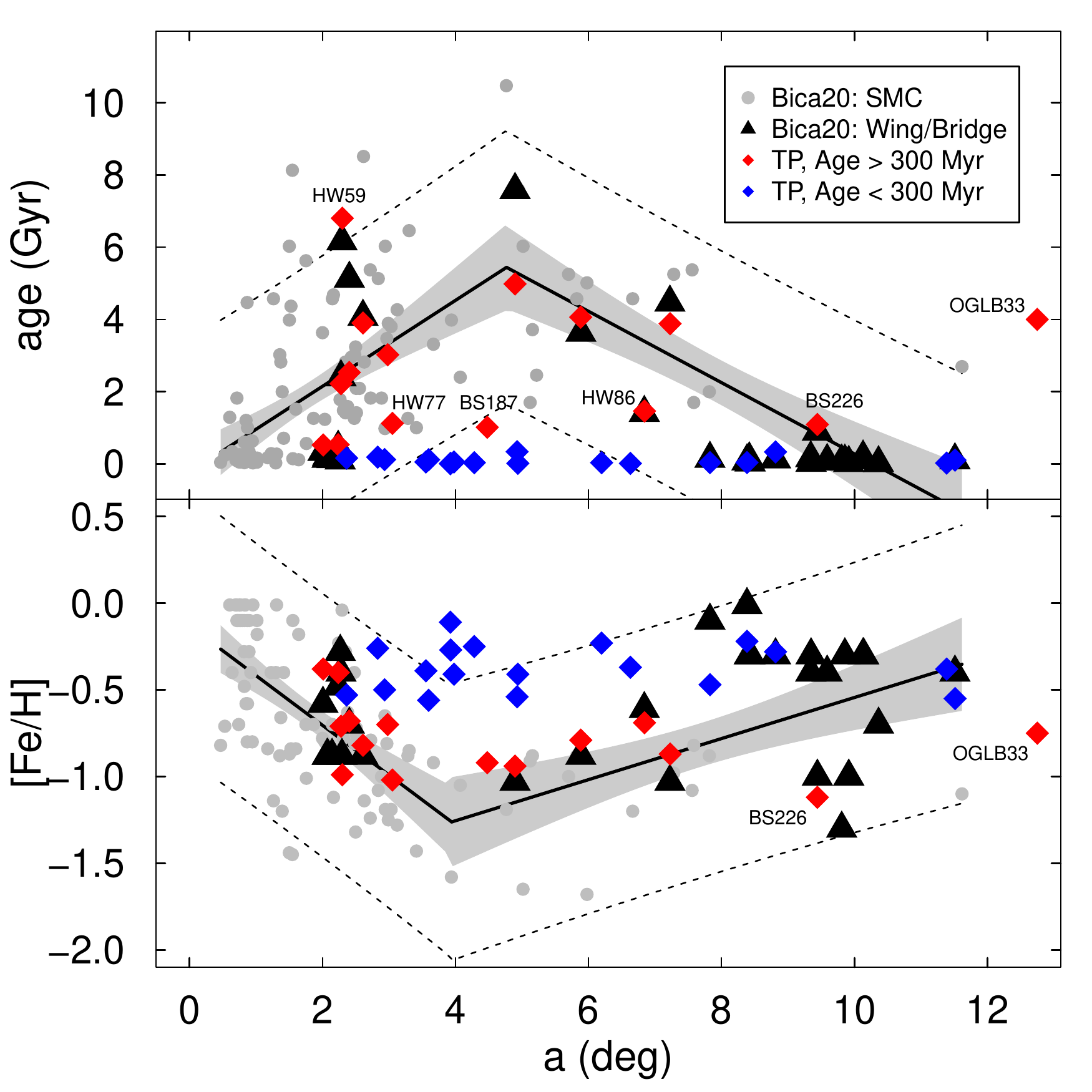}
    \caption{Age and metallicity as a function of the projected distance to the SMC centre. The grey and black symbols are from the \citet{2020AJ....159...82B} catalogue (SMC and Wing/Bridge clusters), and the red and blue diamonds are the present sample of old and young clusters. The solid and dashed lines and the grey shaded areas represent the fits from \citet{2020AJ....159...82B} to their age and metallicity distributions.
    The old sample clusters that most deviate from the age and metallicity gradients are annotated.}
    \label{fig:grad}
\end{figure}

\begin{figure*}
    \centering
    \includegraphics[trim={0cm 0.55cm 0cm 0.5cm},clip,width=0.47\textwidth]{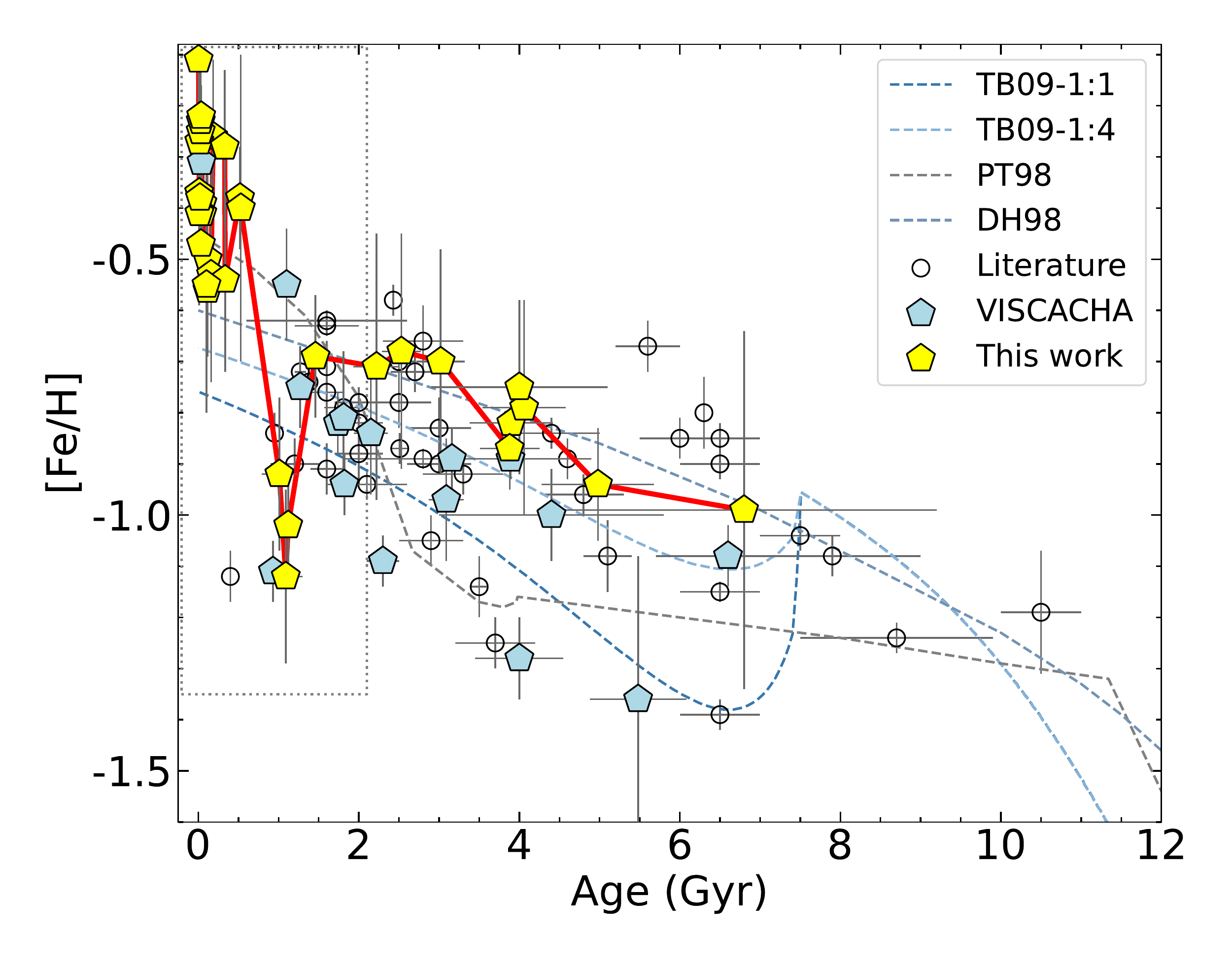}
    \includegraphics[trim={0cm 0.55cm 0cm 0.5cm},clip,width=0.47\textwidth]{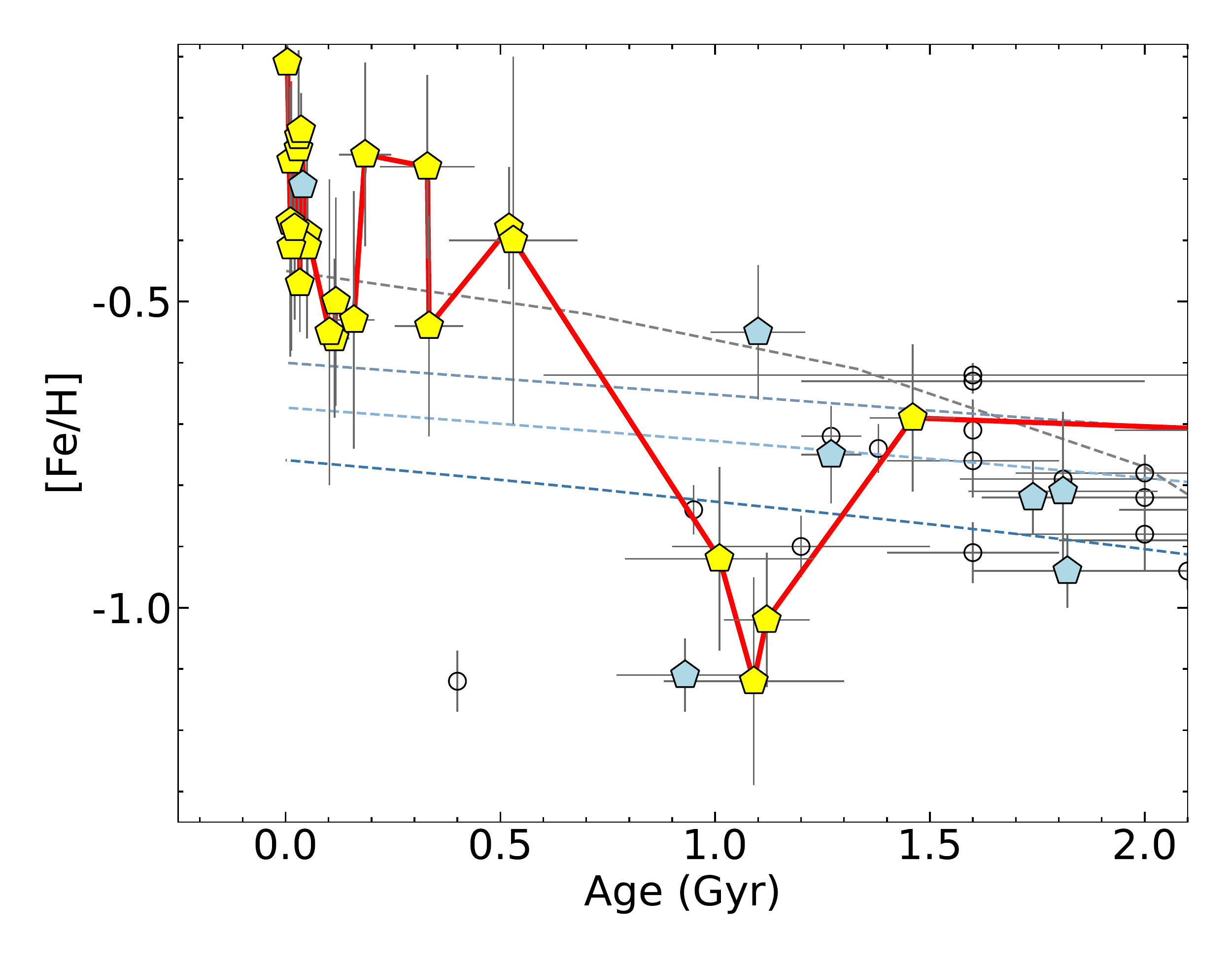}
    \caption{\textit{(Left:)} Age-metallicity relation including the present results (yellow pentagons), previous results from the VISCACHA survey (blue pentagons), and literature data with CaT metallicities (open black circles). Chemical evolution models are overplotted: \citet{1998MNRAS.299..535P},
    \citet{1998AJ....115.1934D} and \citet{2009ApJ...700L..69T}. A red line connects the results, from the older to the younger cluster, suggesting the existence of a large dip around $1-1.5$\,Gyr and smaller one around $200$\,Myr. \textit{(Right:)} Zoom-in of the young counterpart of the age-metallicity relation.
    }
    \label{fig:amr}
\end{figure*}

\section{Discussion}
\label{sec5}

We analysed 33 
stellar clusters located from the SMC Wing until halfway the Bridge towards the LMC,
noting
that farther away
most of the objects are stellar associations \citep{2015MNRAS.453.3190B, 2020AJ....159...82B}. The aim of this work was to verify if there is a metallicity and/or age gradient along the Bridge, between the SMC and the LMC, and to characterize the young and old
cluster
populations in terms of distance and space distribution. 

Figure~\ref{fig:spatial-ages} shows the projected distribution of the sample clusters, overplotted on the 449 Bridge clusters and associations. 
The grey points are all clusters and associations reported in \citet{2020AJ....159...82B}, the black diamonds are
the ones classified as Bridge objects
and the colour-coded circles are the sample clusters according to their age 
or metallicity.
Most of the older clusters are located close to the SMC,
whereas beyond the SMC Wing the number of old and young clusters are comparable (with a low number statistics).
It suggests a stratification of the cluster groups of similar age, with more recent cluster formation predominating in Bridge regions.

Two groups of clusters are clearly detected in our results:
13 Bridge clusters more metal-poor than $\rm{[Fe/H]}=-0.6$\,dex
(more consistent with the SMC metallicity) and older than $500$\,Myr; and 15 young and metal-rich clusters ($-0.5\lesssim \rm{[Fe/H]\:(dex)}<-0.1$), probably formed in-situ after the recent encounter $150-250$\,Myr ago.
Such high metallicities indicate that the gas used to form clusters
was originated possibly from a more metal-rich component in the SMC centre \citep{2018MNRAS.478.5017R} or from the LMC.
The ages of the remaining five clusters ($200-500$\,Myr, namely B147, WG13, B165, K55 and K57) with metallicity between $\rm{[Fe/H]}\sim-0.6$ and $-0.4$\,dex indicate they formed just before the Bridge and require further analysis.
According to \citet{2012MNRAS.421.2109B}, their Model 2 predicts that most of the Bridge material came from the SMC with some smaller contribution from the LMC, in agreement with the observations.
Based on these two groups, hereafter we make a distinction between the clusters younger and older than $300$\,Myr (formed in situ vs. stripped), in order to check their relation with the HI gas, their spatial distribution and gradients.

Figure~\ref{fig:hi-map-ages} presents a map of neutral hydrogen (HI) column density from the HI4PI survey \citep{2016A&A...594A.116H} along the SMC eastern side and the Bridge. All the sample clusters are concentrated in regions with higher gas density; however, if we consider only the clusters farther from the SMC Wing ($\rm{RA} > 1^h20^m$), the young ones 
appear
to be located in regions with slightly higher density on average, compared to the older ones.
A further look into the kinematics of the two groups of clusters is needed to provide a stronger conclusion about their association to the HI gas. However, since \textit{Gaia} proper motions are available only for the very bright stars of these clusters and no radial velocities are available, we leave this analysis for a future work.

We converted the RA, DEC coordinates combined with the line-of-sight distances derived in this work into a Cartesian system centred at the SMC centre following the equations by \citet{2001AJ....122.1807V}. The $z=0$ plane is tangent to the sky at $(x,y)=(0,0)$, where $z$ increases towards the observer, $x$ increases towards West, $y$ increases
towards North.
Figure~\ref{fig:mm0vsRA} shows the projections of the distribution of the sample clusters around the SMC, of which four are within the SMC tidal radius of 4\,kpc (\citetalias{2022MNRAS.512.4334D}). The young clusters appear to follow a homogeneous distribution up to a radius of 13\,kpc from the SMC centre, whereas the old ones are gathered in some specific regions. The fact that most of the sample clusters are at distances smaller than the SMC ($z>0$) and pointing to the LMC seems consistent with the recent collision scenario \citep[e.g.][]{2022ApJ...927..153C},
possibly
with the SMC moving away after that and leaving gas and both the young and old stellar populations in its path.

In order to check if the Wing/Bridge clusters follow similar age and metallicity trends compared to the SMC clusters, in Figure~\ref{fig:grad}
we overplotted our results over the figure~8 from \citet{2020AJ....159...82B}.
In this figure, the red and blue diamonds are the sample clusters
that are older and younger than $300$\,Myr, respectively. 
The old sample clusters follow the overall age and metallicity
radial distributions
of the SMC
clusters.
The young clusters clearly do not follow the
radial distributions from \citet{2020AJ....159...82B}.
The clusters within
$a<4^\circ$ have ages of $100-300$\,Myr, whereas the
outer
Wing/Bridge clusters have
ages
$<100$\,Myr. The metallicity of these young clusters also deviate from the 
radial distributions from \citet{2020AJ....159...82B},
with an offset to more metal-rich values and a mean value of $\rm{[Fe/H]}\sim-0.4$\,dex.

The peak metallicity of the SMC main body for the younger stellar populations is about $\sim -0.5$\,dex 
\citep[e.g.][]{2018MNRAS.478.5017R}. We interpret our results as evidence that the gas that was stripped out from the SMC to form the Magellanic Bridge
came from the innermost region of the SMC that was enriched by stellar evolution. This metal-rich gas was the material used to form the star clusters along the whole extension of the gaseous
Wing
and beginning of the Bridge ($\rm{RA} < 3^h$), forming clusters with similar metallicity around $\rm{[Fe/H]} = -0.4$\,dex. This result is in agreement with the analysis of B-type supergiants \citep{2005A&A...429.1025L}. The more distant gaseous bridge ($\rm{RA} > 3^h$) seems to be more metal-poor \citep[e.g.][]{2021A&A...646A..16R}, which is beyond the scope of this work, nevertheless it is an evidence that the formation of the whole extension of the Bridge
has
a complex history.
In this sense, we speculate that the clusters older than the formation of the Bridge are more metal-poor and were dragged from the SMC along with the gas during the formation of the Bridge.

In Figure~\ref{fig:amr} we show the age-metallicity relation (AMR) including literature and present results.
Based on the study by \citet[in particular the AMR in their figure~16]{2022A&A...662A..75P},
in the present AMR plot we selected
three popular chemical evolution models,
namely the classical \citet[PT98]{1998MNRAS.299..535P},
the closed box model from \citet[DH98]{1998AJ....115.1934D}
and the two merger scenarios by \citet[TB09]{2009ApJ...700L..69T}.
The latter includes 
mergers with a mass ratio of 1:1 and 1:4
that should have occurred
$\sim7.5$\,Gyr ago.
In this Figure, black circles and blue pentagons have spectroscopic metallicities
from CaT
\citep{1998AJ....115.1934D, 2009AJ....138..517P, 2015AJ....149..154P, 2022A&A...662A..75P, 2022A&A...664A.168D},
whereas the blue pentagons are from the VISCACHA collaboration (\citetalias{2019MNRAS.484.5702M}, \citetalias{2021A&A...647L...9D} and \citetalias{2022MNRAS.512.4334D}).
The three sets of chemical evolution models, despite their significant differences in terms of assumptions, reproduce well the data.

When comparing Figure~\ref{fig:amr} with the AMR from \citet{2022A&A...662A..75P}, it is clear that our data help to better visualise the AMR, in particular including several young objects below 1\,Gyr (which do not have red giants,
and consequently no
measurable CaT lines) and a group of 4 clusters with $\sim 1$\,Gyr and $\rm{[Fe/H]}\sim-1.0$\,dex. Additionally, \citet{2022A&A...662A..75P} isolate in their figure~18 an AMR for three Wing/Bridge clusters (L110, HW86 and L113), and we complement the sample with ten more clusters older than $1$\,Gyr.

A few important remarks can be deduced from the present AMR plot:
\textit{(i)} a large dip is identified by five metal-poor clusters with ages around $1-1.5$\,Gyr, with the metallicity decreasing from $\rm{[Fe/H]}=-0.6$ to $\sim-1.0$\,dex, followed by a rapid chemical enrichment; \textit{(ii)} a smaller dip is observed to start around $200$\,Myr ago, with the metallicity decreasing from $\rm{[Fe/H]}=-0.3$ to $\sim-0.6$\,dex; \textit{(iii)} the sample clusters older than $1.5$\,Gyr are slightly more metal-rich than the literature and VISCACHA points.
Three clusters at $200$ and $300$\,Myr (B147, WG13 and B165) present some noise in detecting the exact epoch of the metallicity dip, which is consistent with the uncertainties faced by the models in reproducing the recent encounter.

As far as we know, both metallicity dips are detected for the first time in the present work.
Comparing them to the dynamical history of the System, both make sense given the recent encounters between the MCs that formed the Stream and Bridge, $2$\,Gyr and $150-250$\,Myr ago
respectively \citep{2010ApJ...721L..97B, 2016ARA&A..54..363D, 2022ApJ...927..153C}.
The three clusters
that
mark the prominent dip at $1.5$\,Gyr, namely HW77, BS187 and BS198, are the same that deviate from the overall age and metallicity gradients of the SMC. The same applies to two metal-poor clusters from the literature sample: IC1708 \citep[Northern Bridge;][]{2021A&A...647L...9D} and K9 \citep[West Halo;][]{2015AJ....149..154P, 2022A&A...662A..75P}.
Additionally, the young Bridge clusters BS233 and BS235 were identified to have $\rm{[Fe/H]}=-1.3$ and $-1.0$\,dex by \citet{2015MNRAS.453.3190B, 2020AJ....159...82B} from CMDs.

\citet{2007ApJ...658..345H} suggested that the star formation in the Bridge commenced about $200-300$\,Myr ago and continued over an extended interval, until about 40\,Myr ago. They determine the Bridge star formation history over a wide area and, by applying a Kroupa mass function and a $10$\,Gyr stellar population, they calculate an upper mass limit for the Bridge of $1.5\times 10^4 M_\odot$.
In the present analysis of around one third of the Bridge clusters, we obtain a total mass of $10^5 M_\odot$. Extrapolating this value to the $\sim 100$ clusters and $\sim 300$ associations in the Wing/Bridge \citep{2020AJ....159...82B}, 
a conservative estimate of $3-5\times 10^5 M_\odot$ appears to be more reasonable
for the Bridge stellar mass.

\citet{2018ApJ...864...55Z} ran models to explain kinematic information and concluded that the Bridge was formed about $150-250$\,Myr ago. As a consequence, the young stellar population would have been formed in-situ in the Bridge, therefore representing the nearest extragalactic stellar population formed from tidal debris. The fact the Wing/Bridge older clusters are more metal-rich than the overall SMC cluster distribution could indicate that they were dragged from the SMC main body that is more metal-rich than the SMC outskirts \citep{2010A&A...517A..50G,2018MNRAS.478.5017R}.

\section{Conclusions}
\label{sec6}

We have derived ages, metallicities, distances and masses for 33 
star clusters from deep observations obtained with the adaptive optics of the SAMI imager, at the SOAR telescope. These clusters are among the $\sim 100$ star clusters located in the Bridge between the SMC and LMC. The sample includes a few clusters in the SMC Wing and along the Bridge ($\rm{RA} < 3^h$) but not close to the LMC, because between the LMC and halfway the Bridge there are essentially no clusters but only sparse stellar associations. Our results include metallicities for $18$ clusters and ages for $9$ clusters derived for the first time. Based on the masses of our sample clusters we estimate a minimum stellar mass of the Bridge to be $3-5\times 10^5\,M_\odot$, more than one magnitude higher than previous estimates.

Interestingly, we have found 13 metal-poor clusters ($\rm{[Fe/H]}<-0.6$\,dex) with ages between 500\,Myr and 6.8\,Gyr. They were probably formed from SMC gas and stripped from the SMC, as predicted by \citet[Model 2]{2012MNRAS.421.2109B}. We showed that these clusters follow strictly the SMC age and metallicity gradients, as well as age-metallicity relation given by the chemical evolution models.

We have also found 15 young ($\lesssim 200$\,Myr) and metal-rich ($-0.5<\rm{[Fe/H]\:(dex)}<-0.1$) clusters, which probably formed in-situ after the Bridge formation. Given the more metal-rich regime, the gas that originated this population
could have been extracted from the more metal-rich
component in the SMC centre \citep{2018MNRAS.478.5017R}. This population does not follow the SMC age and metallicity gradients,
is roughly constant around $\rm{[Fe/H]} = -0.4$\,dex, which indicates that they were formed in situ from the metal-rich gas that constitute the Wing and Bridge closer to the SMC ($\rm{RA} < 3^h$). An extraction of gas from the LMC is less likely because the Bridge region closer to the LMC is actually more metal-poor.

A particularly interesting group are the intermediate-age ($\sim 1$\,Gyr) and metal-poor ($\rm{[Fe/H]}<-0.8$\,dex) Bridge clusters, namely BS187, HW77 and 
BS226, together with BS233 and BS235 \citep{2015MNRAS.453.3190B}, and another two from \citet{2022A&A...662A..75P} and \citet{2022A&A...664A.168D}: IC1708 (Northern Bridge) and K9 (West Halo). These clusters mark a metallicity dip on the age-metallicity relation around $1-1.5$\,Gyr, decreasing the metallicity from $-0.6$ to $-1.0$\,dex. It is possible that this episode, followed by a rapid chemical enrichment, happened due to the infall of metal-poor gas, which is consistent with the formation epoch of the Magellanic Stream. These clusters are good candidates to be observed with high-resolution spectroscopy with the future giant telescopes.

Another metallicity dip seems to be present at around $200$\,Myr when metallicity dropped from $-0.25$ to $-0.55$\,dex approximately,
suggesting a second infall of metal-poor gas.
This was the time when the Magellanic Bridge itself was formed \citep{2018ApJ...864...55Z}. In summary, the formation of the Magellanic Stream and Magellanic Bridge left marks in the chemical evolution of the SMC 
and Wing/Bridge
clusters. Dedicated chemical evolution models shall enlighten the explanation of these metallicity dips.

\section*{Acknowledgements}

We thank an anonymous referee for the remarks that improved this paper.
This study was financed in part by the Coordena\c c\~ao de Aperfei\c coamento de Pessoal de N\'ivel Superior - Brasil (CAPES) - Finance Code 001. R.A.P.O. acknowledges FAPESP (proc. 2018/22181-0). F.F.S.M. acknowledges financial support from Conselho Nacional de Desenvolvimento Científico e Tecnológico - CNPq (proc. 404482/2021-0) and from FAPERJ (proc. E-26/201.386/2022 and E-26/211.475/2021). B.B. and E.B. acknowledges partial financial support from FAPESP, CNPq and CAPES.
B.D. acknowledges support by ANID-FONDECYT iniciación grant No. 11221366. J.F.C.S. acknowledges support by FAPEMIG (proc. OET-00020-22). S.O.S. acknowledges FAPESP (proc. 2018/22044-3), the support of Deutsche Forschungsgemeinschaft (DFG, project number 428473034), and the DGAPA–PAPIIT grant IA103122.
L.O.K. acknowledges partial financial support by CNPq (proc. 313843/2021-0) and UESC (proc. 073.6766.2019.0013905-48).
D.M. gratefully acknowledges support by the ANID BASAL projects ACE210002 and FB210003 and by Fondecyt Project No. 1220724. A.P.-V. acknowledges the DGAPA-PAPIIT grant IA103122.

Based on observations obtained at the Southern Astrophysical Research (SOAR) telescope (projects SO2016B-018, SO2017B-014, CN2018B-012, SO2019B-019, SO2020B-019, SO2021B-017), which is a joint project of the Ministério da Ciência, Tecnologia, e Inovação (MCTI) da República Federativa do Brasil, the U.S. National Optical Astronomy Observatory (NOAO), the University of North Carolina at Chapel Hill (UNC), and Michigan State University (MSU).

This work has made use of data from the European Space Agency (ESA) mission {\it Gaia} (\url{https://www.cosmos.esa.int/gaia}), processed by the {\it Gaia} Data Processing and Analysis Consortium (DPAC, \url{https://www.cosmos.esa.int/web/gaia/dpac/consortium}). Funding for the DPAC has been provided by national institutions, in particular the institutions participating in the {\it Gaia} Multilateral Agreement.

\section*{Data availability}

The data underlying this article are available in the NOIRLab Astro Data Archive (\url{https://astroarchive.noirlab.edu/}) or upon
request to the authors.




\bibliographystyle{mnras}
\bibliography{bibliog} 



\appendix

\section{CMDs with the isochrone fits}
\label{app:CMDs}

\begin{figure*} 
    \centering
    \includegraphics[trim={0 0.4cm 2.55cm 0},clip, height=7.7cm]{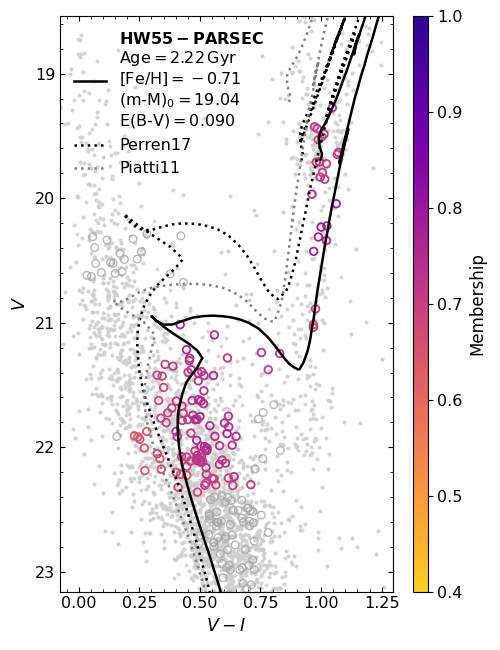}
    \includegraphics[trim={0 0.4cm 2.55cm 0},clip, height=7.7cm]{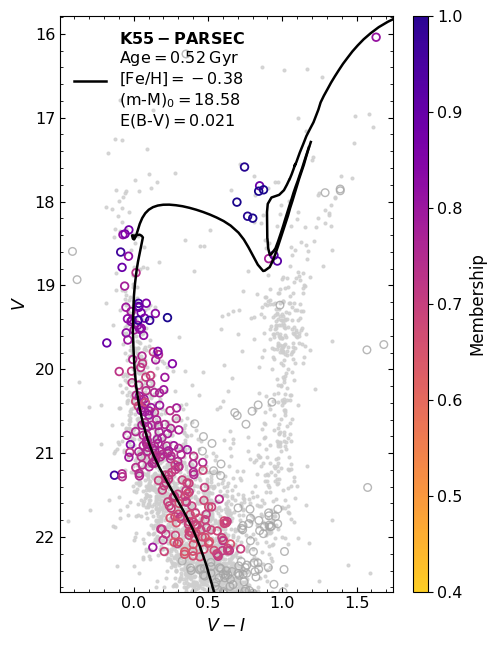}
    \includegraphics[trim={0 0.4cm 0 0},clip, height=7.7cm]{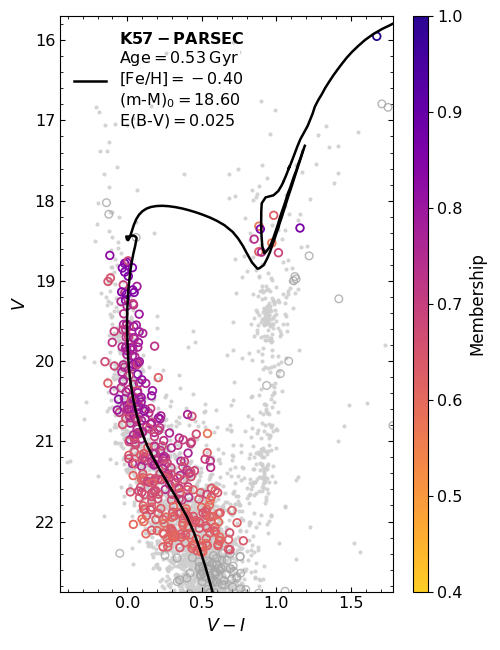}
    
    \includegraphics[trim={0 0.4cm 2.55cm 0},clip, height=7.7cm]{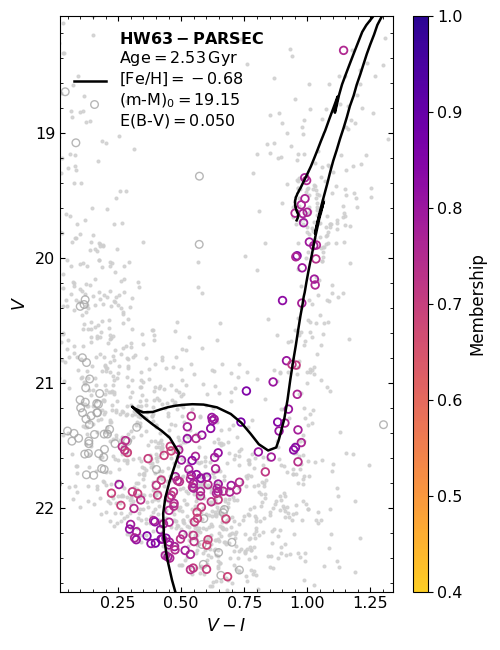}
    \includegraphics[trim={0 0.4cm 2.55cm 0},clip, height=7.7cm]{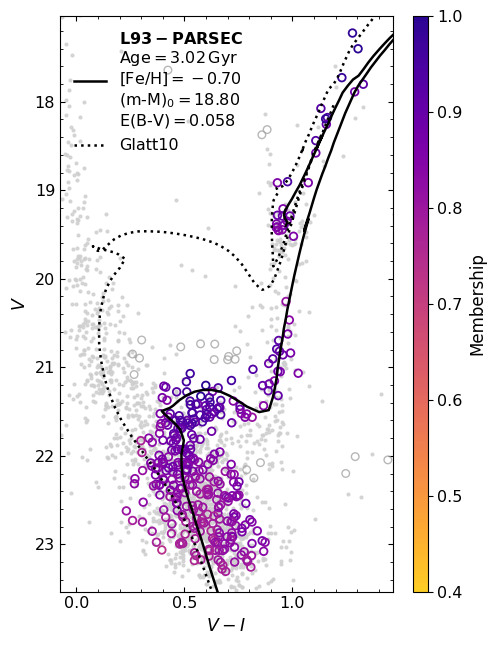}
    \includegraphics[trim={0 0.4cm 0 0},clip, height=7.7cm]{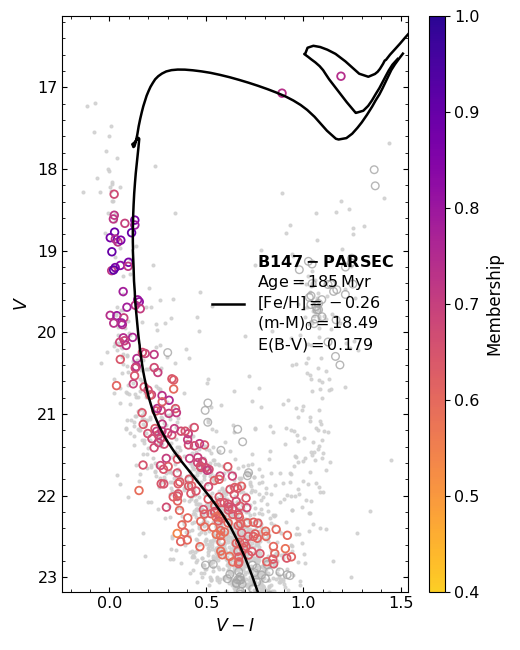}
    
    \includegraphics[trim={0 0.4cm 2.55cm 0},clip, height=7.7cm]{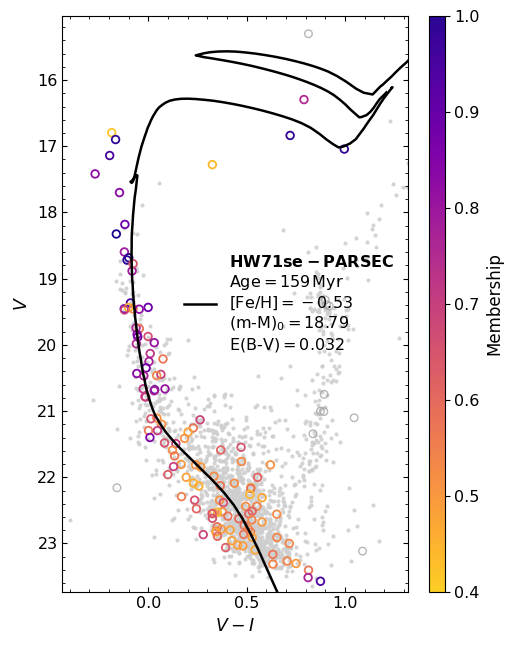}
    \includegraphics[trim={0 0.4cm 2.55cm 0},clip, height=7.7cm]{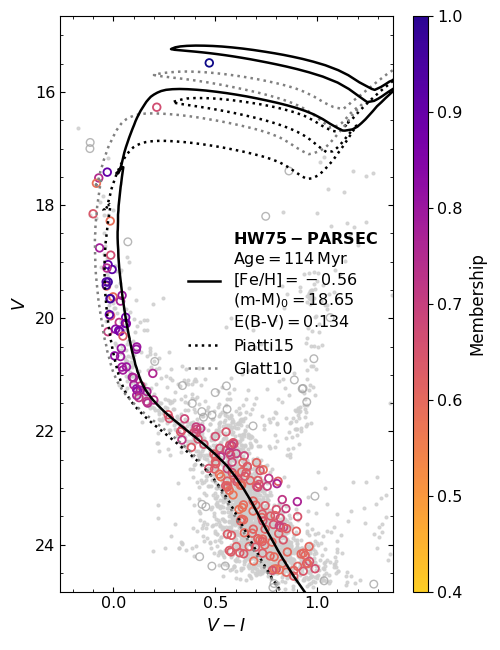}
    \includegraphics[trim={0 0.4cm 0 0},clip, height=7.7cm]{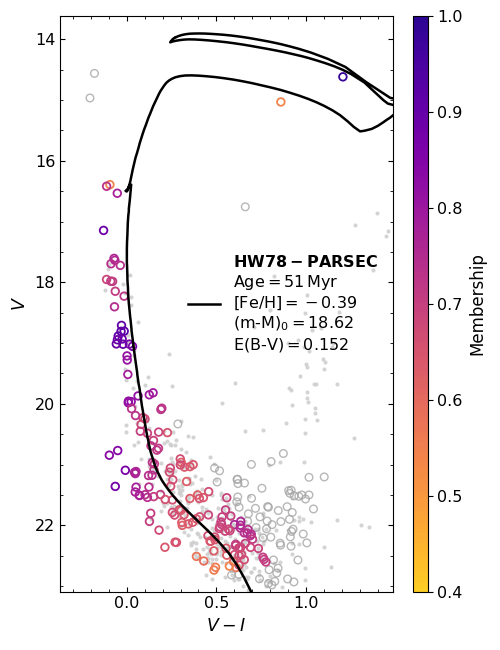}
    \caption{Decontaminated $V$ vs. $V-I$ CMDs containing the results for HW55, K55, K57, HW63, L93, B147, HW71se, HW75 and HW78.}
    \label{fig:appendix1}
\end{figure*}
\begin{figure*} 
    \centering
    \includegraphics[trim={0 0.4cm 2.55cm 0},clip, height=7.7cm]{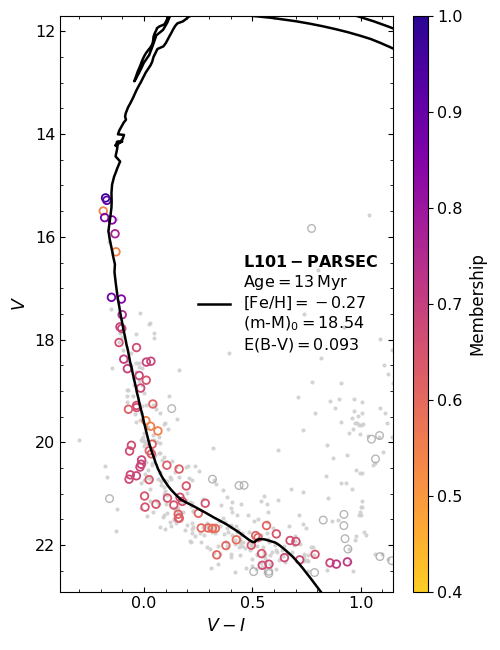}
    \includegraphics[trim={0 0.4cm 2.55cm 0},clip, height=7.7cm]{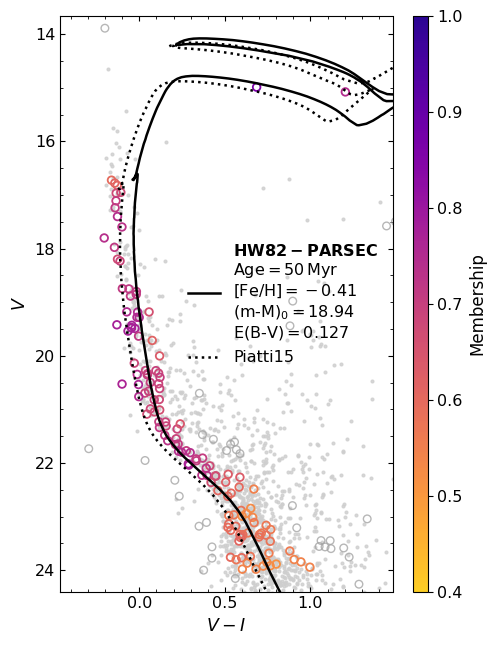}
    \includegraphics[trim={0 0.4cm 0 0},clip, height=7.7cm]{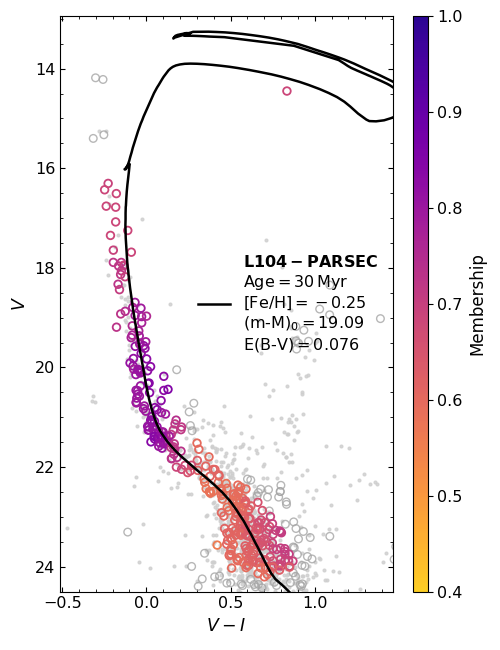}
    
    \includegraphics[trim={0 0.4cm 2.55cm 0},clip, height=7.7cm]{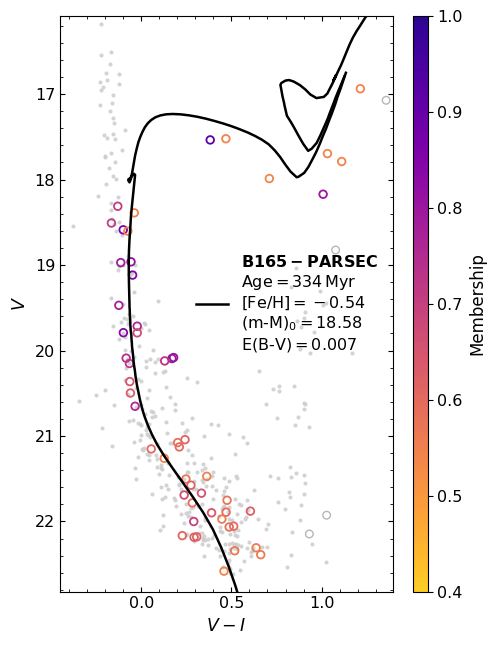}
    \includegraphics[trim={0 0.4cm 2.55cm 0},clip, height=7.7cm]{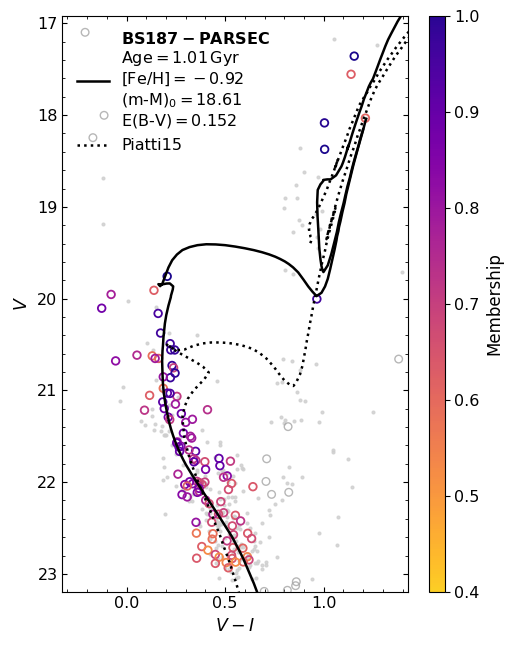}
    \includegraphics[trim={0 0.4cm 0 0},clip, height=7.7cm]{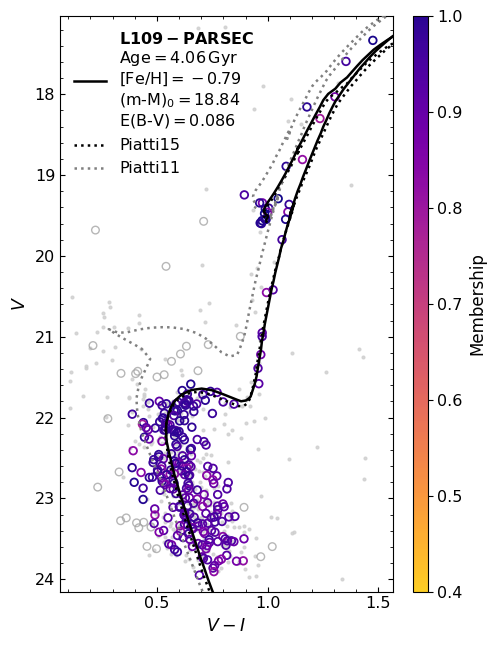}
    
    \includegraphics[trim={0 0.4cm 2.55cm 0},clip, height=7.7cm]{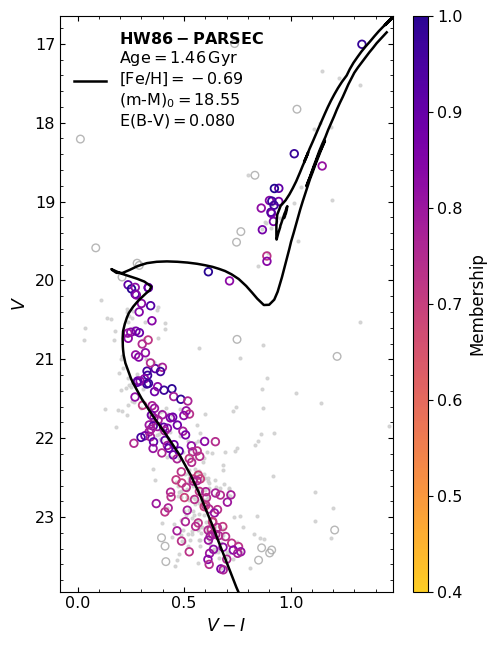}
    \includegraphics[trim={0 0.4cm 2.55cm 0},clip, height=7.7cm]{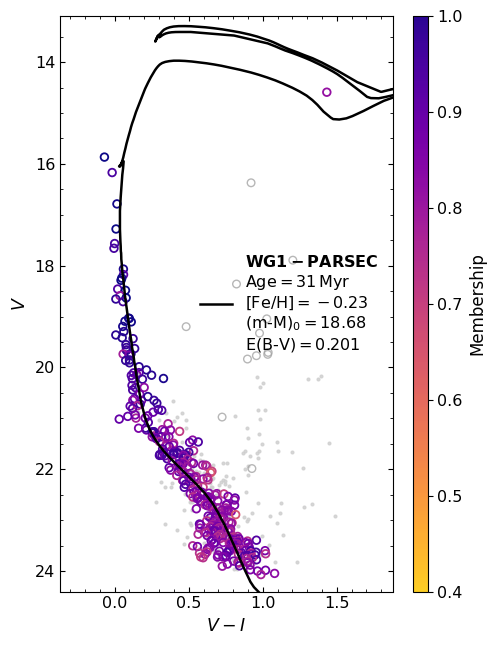}
    \includegraphics[trim={0 0.4cm 0 0},clip, height=7.7cm]{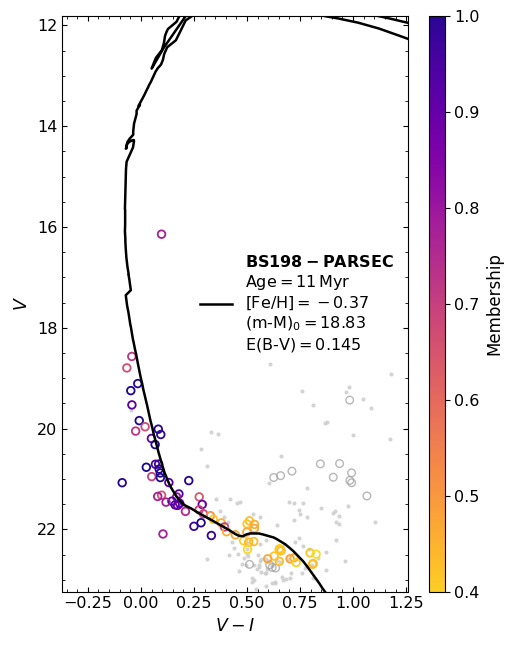}
    \caption{Same as Figure~\ref{fig:appendix1}, with the results for L101, HW82, L104, B165, BS187, L109, HW86, WG1 and BS198.}
    \label{fig:appendix2}
\end{figure*}
\begin{figure*} 
    \centering
    \includegraphics[trim={0 0.4cm 2.55cm 0},clip, height=7.7cm]{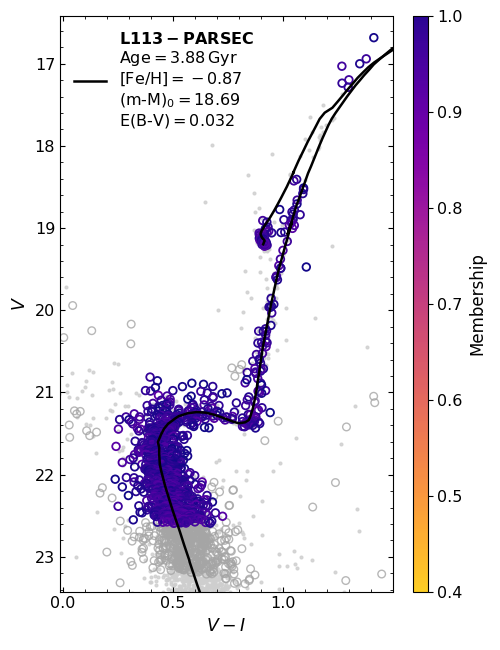}
    \includegraphics[trim={0 0.4cm 2.55cm 0},clip, height=7.7cm]{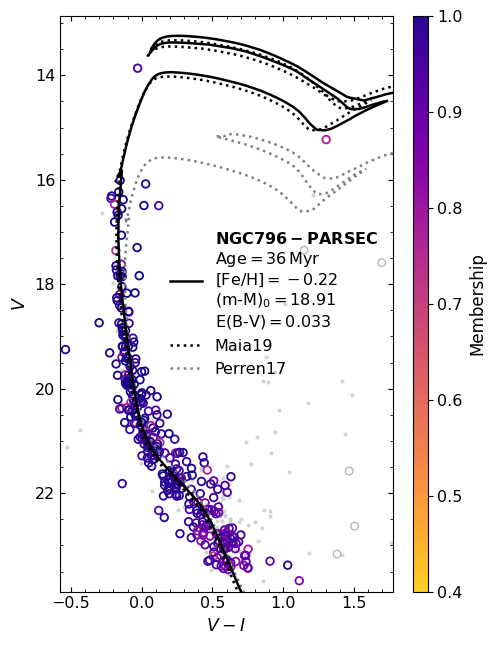}
    \includegraphics[trim={0 0.4cm 0 0},clip, height=7.7cm]{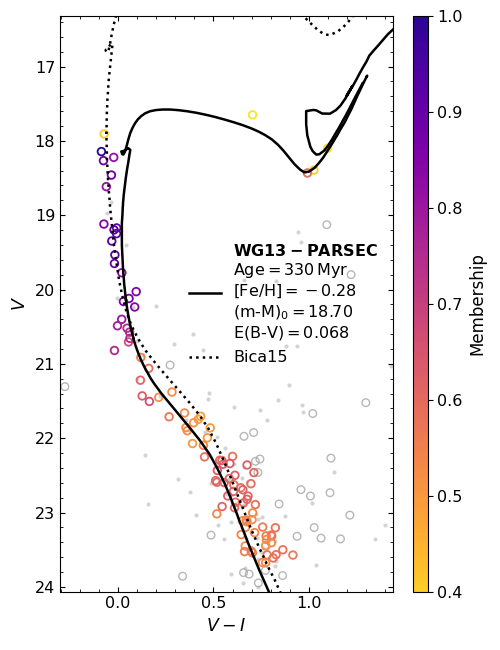}
    
    \includegraphics[trim={0 0.4cm 2.55cm 0},clip, height=7.7cm]{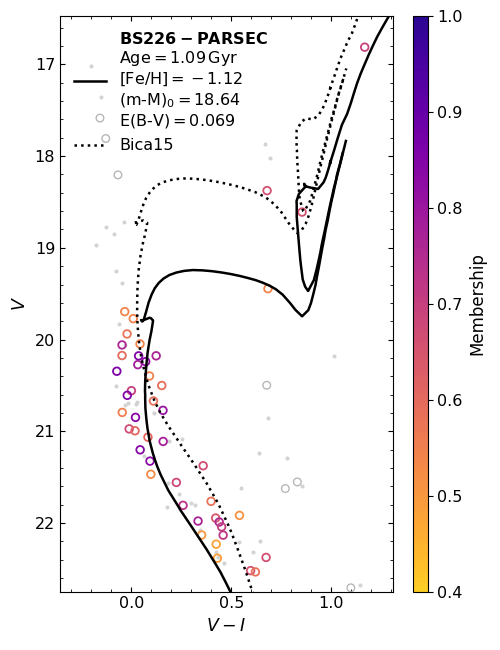}
    \includegraphics[trim={0 0.4cm 2.55cm 0},clip, height=7.7cm]{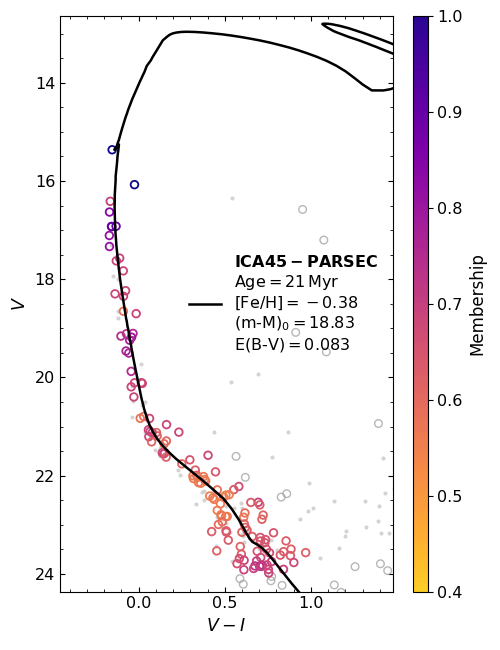}
    \includegraphics[trim={0 0.4cm 0 0},clip, height=7.7cm]{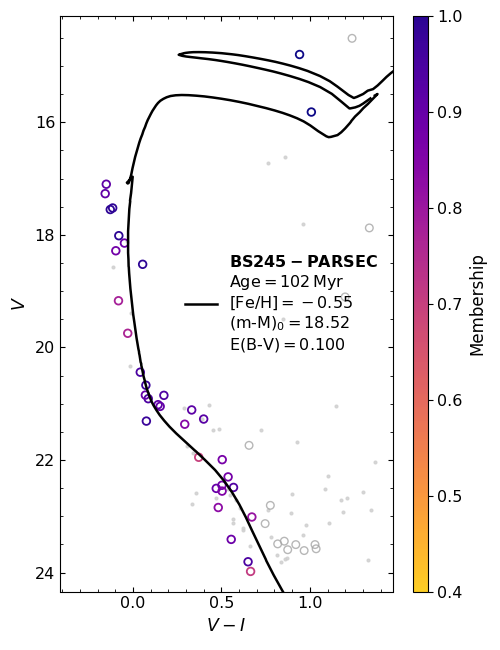}
    
    \includegraphics[trim={0 0.4cm 2.55cm 0},clip, height=7.7cm]{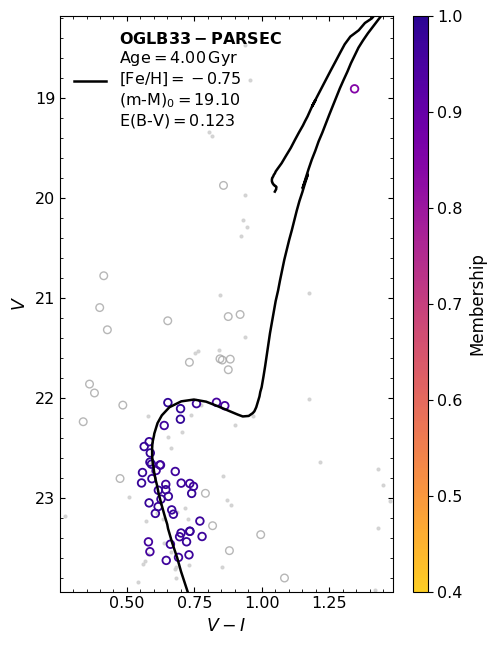}
    \caption{Same as Figure~\ref{fig:appendix1}, with the results for L113, NGC796, WG13, BS226, ICA45, BS245 and OGLB33.}
    \label{fig:appendix3}
\end{figure*}

Figures~\ref{fig:appendix1}, \ref{fig:appendix2} and \ref{fig:appendix3} present the decontaminated $V$ vs. $V-I$ CMDs of the remaining Bridge clusters. The CMDs are colour-coded by membership probability and include the best-fitting isochrone, as well as previous literature results when available. 

\bsp	
\label{lastpage}
\end{document}